\documentclass[12pt,fleqn,english]{extarticle}
\usepackage[T1]{fontenc}
\usepackage{geometry}
\usepackage{color}
\usepackage{babel}
\usepackage{float}
\usepackage{amsmath}
\usepackage{amsthm}
\usepackage{amssymb}
\usepackage{setspace}
\usepackage[unicode=true,pdfusetitle,
 bookmarks=true,bookmarksnumbered=false,bookmarksopen=false,
 breaklinks=false,pdfborder={0 0 1},backref=false,colorlinks=false]
 {hyperref}

\makeatletter

\providecommand{\tabularnewline}{\\}

\theoremstyle{plain}
\newtheorem{thm}{\protect\theoremname}
\theoremstyle{definition}
\newtheorem{defn}[thm]{\protect\definitionname}
\theoremstyle{plain}
\newtheorem{lem}[thm]{\protect\lemmaname}
\theoremstyle{plain}
\newtheorem{assumption}[thm]{\protect\assumptionname}
\theoremstyle{plain}
\newtheorem{prop}[thm]{\protect\propositionname}
\theoremstyle{plain}
\newtheorem{lyxalgorithm}[thm]{\protect\algorithmname}

\usepackage{babel}

\makeatother

\providecommand{\algorithmname}{Algorithm}
\providecommand{\assumptionname}{Assumption}
\providecommand{\definitionname}{Definition}
\providecommand{\lemmaname}{Lemma}
\providecommand{\propositionname}{Proposition}
\providecommand{\theoremname}{Theorem}

\begin{document}
\title{Spanning analysis of stock market anomalies under Prospect Stochastic
Dominance}
\author{Stelios Arvanitis, Olivier Scaillet, Nikolas Topaloglou}
\date{This version: April 2020 }
\maketitle
\begin{abstract}
We develop and implement methods for determining whether introducing
new securities or relaxing investment constraints improves the investment
opportunity set for prospect investors. We formulate a new testing procedure
for prospect spanning for two nested portfolio sets based on subsampling
and Linear Programming. In an application, we use the prospect spanning
framework to evaluate whether well-known anomalies are spanned by
standard factors. We find that of the strategies considered, many
expand the opportunity set of the prospect type investors, thus have
real economic value for them. In-sample and out-of-sample results
prove remarkably consistent in identifying genuine anomalies for prospect
investors.
\end{abstract}

\section{Introduction}

Traditional models in economics and finance assume that investors
evaluate portfolios according to the expected utility framework. The
theoretical motivation for this goes back to Von Neumann and Morgenstern
(1944). Nevertheles, experimental and empirical work has shown that
people systematically violate Expected Utility theory when choosing
among risky assets. Prospect theory, first described by Kahneman and
Tversky (1979) (see also Tversky and Kahneman (1992)), is widely viewed as a better description of
how people evaluate risk in experimental settings. While the theory
contains many remarkable insights, it has proven challenging to apply
these insights in asset pricing, and it is only recently that there
has been real progress in doing so (Barberis et al.\ (2019)). Barberis
and Thaler (2003) and Barberis (2013) are excellent reviews on behavioral
finance and prospect theory.

Stock market anomalies are key drivers of innovation in asset pricing.
These are tradable portfolio strategies, usually constructed as long-short
portfolios based on the top and bottom deciles of sorted stocks,
according to specific characteristics (anomalies). Under the standard
Mean-Variance (MV) paradigm, establishing a cross-sectional return
pattern as an anomaly involves testing for pricing based on a factor
model. If factors are traded, spanning regressions relate to MV criterion.
Arbitrage pricing stipulates that a portfolio of factors is MV-efficient
and no other portfolio can achieve a higher Sharpe Ratio (SR). In
that sense, an anomaly is a strategy that exhibits higher SR and should
be traded away. However, we can question MV spanning for portfolio
selection if returns do not follow elliptical distributions, or investor
preferences depend on more than the first two moments of the return
distribution. Moreover, experimental evidence (Baucells and Heukamp (2006)) 
suggests that investors
do not always act as risk averters. Instead, under certain circumstances,
they behave in a much more complex fashion, exhibiting characteristics
of both risk-loving and risk-averting. They behave differently on
gains and losses, and they are more sensitive to losses than to gains
(loss aversion). The relevant utility function could be concave for
gains and convex for losses (S-Shaped).

The present study contributes to this literature by introducing, operationalizing
and applying prospect spanning tests for portfolio analysis. The general
research question is whether a given investment possibility set $\mathbb{K}$, namely 
the benchmark set, contains portfolios which prospect dominates
all alternatives in an expanded investment possibility set $\mathbb{L}$.

Stochastic spanning (Arvanitis et al.\ (2019)) is a model-free alternative
to MV spanning of Huberman and Kandel (1987) (see also
Jobson and Korkie (1989), De Roon, Neyman, and Werker (2001)). Spanning
occurs if introducing new securities or relaxing investment constraints
does not improve the investment possibility set for a given class
of investors. MV spanning checks if the mean-variance
frontier of a set of assets is identical to the mean-variance frontier
of a larger set made of those assets plus additional assets (Kan
and Zhou (2012), Penaranda and Sentana (2012)). Here we investigate
such a problem for investors with prospect type preferences which
are interested in the whole return distributions generated by two
sets of assets, namely stochastic dominance. First, we introduce the
concept of prospect spanning, which is consistent with prospect type
investors. We propose a theoretical measure for prospect spanning
based on stochastic dominance and derive the exact limit distribution
for the associated empirical test statistic for a general class of
dynamic processes. To check prospect spanning on data, we develop
consistent and feasible test procedures based on subsampling and Linear
Programming (LP).

Similarly to Arvanitis et al.\ (2019), it is easy to see that if the
prospect efficient set is non-empty, a prospect spanning set is essentially
any superset of the former. As such, we can use a prospect spanning
set to provide an outer approximation of the efficient set. This is
useful in at least two ways. First, if the spanning set is small enough,
the problem of optimal choice is reduced to a potentially simpler
problem. Indeed, a spanning set is a reduction of the original portfolio
set without loss of investment opportunities for any investor with
S-shaped preferences. Secondly, if an algorithm for the choice of
non-trivial canditate spanning sets is available, we can use this
to construct decreasing sequences of prospect spanning sets that appropriately
converge to the efficient set. Given the complexity of the prospect
efficient set (see for example Ingersoll (2016)) such an approach
can be useful for the determination of its properties.

The second contribution of the paper is to examine if we can explain well-known stock
market anomalies by standard factor models for prospect
investors. To do so, we test if trading strategies are genuine violations
of standard factor models. More precisely, in the in-sample analysis,
we use the prospect spanning test in order to check  whether a portfolio
set originating from a standard factor model, $\mathbb{K}$, spans
the same set augmented with a market anomaly, $\mathbb{L}$. This check
could be of significant relevance to the empirical analysis of financial
markets. If the hypothesis of prospect spanning holds, the particular
market anomaly can be explained by the factor model. Then the trading strategy that is identified in the literature
as market anomaly may not be an attractive investment opportunity
for prospect investors. On the contrary, if the hypothesis is not
true, then the anomaly expands the opportunity set
for prospect investors, and is useful to that extent. We also examine whether the cross-sectional
patterns that found to expand the set of factors in-sample, maintain
this abnormal return out-of-sample. Therefore, we use out-of-sample
backtesting experiments as an independent criterion for robustness
of in-sample test results (Harvey et al.\ (2016)). It turns out that
prospect spanning tests produce remarkably consistent results both
in- and out-of-sample in identifying trading strategies as genuine
market anomalies for prospect investors. Thus, our framework helps
validating stock market anomalies for prospect preferences.

Benartzi and Thaler (1995) utilize prospect theory to present an approach
called myopic loss aversion which consists of two behavioural concepts,
namely loss aversion and mental accounting. Barberis et al.\ (2001)
study asset prices in an economy where investors derive direct utility
not only from consumption but also from fluctuations in the value
of their financial wealth. They are loss averse over these fluctuations
and how loss averse they are depends on their prior investment performance.
The design of their model is influenced by prospect theory. Barberis
and Huang (2008) study the pricing of financial securities when investors
make decisions according to cumulative prospect theory. Several other
papers confirm that positively skewed stocks have lower average returns
(Boyer, Mitton, and Vorkink (2010), Bali, Cakici, and Whitelaw (2011),
Kumar (2009), Conrad, Dittmar, and Ghysels (2013)). Barberis and Xiong
(2009, 2012) and Ingersoll and Jin (2013) show that theoretical investment
models based on S-Shape utility maximisers help to understand the
disposition effect found empirically in many studies (see e.g. Odean
(1988), Grinblatt and Han (2005), Frazzini (2006), Calvet, Campbell,
and Sodini (2009)). Kyle, Ou-Yang, and Xiong (2006)
provide a formal framework to analyze the liquidation decisions of economic agents under prospect theory.
He and Zhou (2011) 
study the impact of prospect theory on optimal risky exposures in portfolio choice
through an analytical treatment.
Ebert and Strack (2015) set up a general version
of prospect theory and prove that probability weighting implies skewness
preference in the small. Barberis et al.\ (2016) test the hypothesis
that, when thinking about allocating money to a stock, investors mentally
represent the stock by the distribution of its past returns and then
evaluate this distribution in the way described by prospect theory.
Moreover, Barberis et al.\ (2019) present a model of asset prices in
which investors evaluate risk according to prospect theory and examine
its ability to explain prominent stock market anomalies. In our
paper, we test whether well-known factor models span the augmented
universe with a prominent stock market anomaly, and if not, whether
the result is supported out-of sample.

The paper is organised as follows. In Section 2, we review the definition
of prospect stochastic dominance relation and we define the relevant
concept of prospect spanning. We provide with a representation
based on a class of S-shaped utility functions without assuming differentiability.
Using an empirical approximation of the latter, we construct a test
for the null hypothesis of spanning based on subsampling. The construction
is based on the limiting null distribution of the test statistic which
has the form of a saddle type point of a relevant Gaussian process.
Under a weak condition on the structure of the parameter
contact sets, we show that the test is asymptotically exact and consistent.
This is weaker than the parameter extreme point comparisons used in
Arvanitis, Scaillet and Topaloglou (2019) to obtain exactness in large samples.

In Section 3, we provide with a numerical approximation
 of the statistic that is based on the utility representation derived before. The
utility functions are univariate, and normalized. We  use a finite
set of increasing  piecewise-linear functions, restricted
to the bounded empirical supports, that are constructed as convex
mixtures of appropriate "ramp functions'' ( in the spirit of Russel and Seo (1989)) in our representation. For
every such utility function, we solve two embedded linear maximization
problems. This is an improvement over the implementation in Arvanitis and Topaloglou
(2017) and Arvanitis, Scaillet and Topaloglou (2019) where they formulate tests in terms of Mixed-Integer Programming (MIP) problems. MIP problems
are NP-complete, and far more difficult to solve. Our numerical approximations
are simple and fast since they are based on standard LP. They suit better
 resampling methods, which otherwise become quickly computationally demanding in empirical work.

In Section 4, we perform an empirical application where we use the prospect
spanning tests to evaluate stock market anomalies using standard factor
models. We consider three such models that build on the pioneer three-factor
model of Fama and French (1993): the four-factor model of Hou, Xue
and Zhang (2015), the five-factor model of Fama and French (2015),
and the four-factor model of Stambaugh and Yuan (2017). Given the
extensive set of results produced under alternative spanning criteria,
the analysis is confined to 11 well-known strategies used to construct
Stambaugh-Yuan factors, along with 7 extra (18 overall) that attracted
significant attention, namely Betting against Beta, Quality minus Junk,
Size, Growth Option, Value (Book to Market), Idiosyncratic Volatility
and Profitability. The 11 anomalies used in Stambaugh and Yuan (2017)
are realigned appropriately to yield positive average returns. In
particular, anomaly variables that relate to investment activity (Asset Growth, Investment to  Assets, Net Stock Issues, Composite Equity Issue,
Accruals) are defined low-minus-high decile portfolio returns,
rather than high-minus-low. All the other anomalies are constructed
as high-minus-low decile portfolio returns. These 18 trading strategies
constitute our playing field for comparing spanning test results.
Yet, we emphasize that this paper is not intended to compare factor models
in terms of their ability to capture the cross-section of expected
returns under prospect preferences. Instead, we use alternative factor models as a robustness check for testing
the consistency of in- and out-of-sample results under the prospect spanning framework.
Each factor
model is our initial system of investment coordinates which we take as a granted opportunity set, without questioning its asset
pricing validity. We view here the factors solely as investable assets (since they correspond to tradable strategies based on asset portfolios), and similarly for the anomalies. The anomalies might be labelled  by other authors as factors if indeed priced in the cross-section, but we do not address such a research question in this paper.

Finally, Section 5 concludes the paper. In Appendix A, we provide a short description of the stock market anomalies used in the empirical application. In Appendix B, we also provide
a short description of the performance measure used in the out-of-sample analysis.  We give in a
 separate Online  Appendix:  i) the limiting properties of the testing procedures
under sequences of local alternatives, ii) a Monte Carlo study of the finite sample properties
of the test, iii) the proofs of the main results, as well as auxiliary lemmata and their proofs,
iv) summary statistics of the factor and anomaly returns over our sample period from January 1974 to December 2016, and v)
additional empirical results on out-of-sample analysis of market anomalies.

\section{Prospect Stochastic Dominance and Stochastic Spanning}

The theory of stochastic dominance (SD) gives a systematic framework
for analyzing investor behavior under uncertainty (see Chapter 4 of
Danthine and Donaldson (2014) for an introduction oriented towards
finance). Stochastic dominance ranks portfolios based on general regularity
conditions for decision making under risk (see Hadar and Russell (1969),
Hanoch and Levy (1969), and Rothschild and Stiglitz (1970)). SD uses
a distribution-free assumption framework which allows for nonparametric
statistical estimation and inference methods. We can see SD as a
flexible model-free alternative to mean-variance dominance of Modern Portfolio
Theory (Markowitz (1952)). The mean-variance criterion is consistent with Expected Utility
for elliptical distributions such as the normal distribution (Chamberlain
(1983), Owen and Rabinovitch (1983), Berk (1997)) but has limited
economic meaning when we cannot completely characterize the probability
distribution by its location and scale. Simaan (1993), Athayde and
Flores (2004), and Mencia and Sentana (2009) develop a mean-variance-skewness
framework based on generalizations of elliptical distributions that
are fully characterized by their first three moments. SD presents
a further generalization that accounts for all moments of the return
distributions without necessarily assuming a particular family of distributions.

Inspired by previous work, Levy and Levy (2002) formulate the notions
of prospect stochastic dominance (PSD) (see also Levy and Wiener (1998),
Levy and Levy (2004)) and Markowitz stochastic dominance (MSD). Those
notions extend the well-know first degree stochastic dominance (FSD)
and second degree stochastic dominance (SSD). PSD and MSD investigates
choices by investors who have S-shaped utility functions and reverse
S-shaped utility functions. Arvanitis and Topaloglou (2017) develop
consistent tests for PSD and MSD efficiency which is an extension
to the case where full diversication is allowed. Arvanitis, Scaillet and Topaloglou (2019) investigate MSD spanning.
This paper extends those works to prospect spanning, which is consistent
with prospect preferences.

\subsection{Stochastic Spanning for Prospect Dominance and Analytical Representation}

Given a probability space $\left(\Omega,\mathcal{F},\mathbb{P}\right)$,
suppose that $F$ denotes the cdf of some probability measure on $\mathbb{R}^{n}$.
Let $G(z,\lambda,F)$ be $\int_{\mathbb{R}^{n}}1_{\{\lambda^{T}u\leq z\}}dF(u)$,
i.e., the cdf of the linear transformation $x\in\mathbb{R}^{n}\rightarrow\lambda^{T}x$
where $\lambda$ assumes its values in $\mathbb{L}$, which denotes
the portfolio space. We suppose that the portfolio space is a closed
non-empty subset of $\mathbb{S}=\{\lambda\in\mathbb{R}_{+}^{n}:$\textbf{
$\boldsymbol{1}^{T}\lambda$}$=1,\}$, possibly formulated by further
economic, legal restrictions, etc. In many applications, we have that
$\mathbb{L}=\mathbb{S}$. We denote by $\mathbb{K}$ a distinguished
subcollection of $\mathbb{L}$ and generic elements of $\mathbb{L}$
by $\lambda,\kappa$, etc. In order to define the concepts of PSD
and subsequently of stochastic spanning, we consider 
$\mathcal{J}(z_{1},z_{2},\lambda;F):=\int_{z_{1}}^{z_{2}}G\left(u,\lambda,F\right)du.
$

\begin{defn}
\label{P-dom}$\kappa$ \emph{weakly} \emph{Prospect-dominates} $\lambda$,
written as $\kappa\succcurlyeq_{P}\lambda$, iff 
we have the inequalities 
$P_{1}\left(z,\lambda,\kappa,F\right):=\mathcal{J}\left(z,0,\kappa,F\right)-\mathcal{J}\left(z,0,\lambda,F\right)\leq0,\:\forall z\in\mathbb{R}_{-}
$
and 
$
P_{2}\left(z,\lambda,\kappa,F\right):=\mathcal{J}\left(0,z,\kappa,F\right)-\mathcal{J}\left(0,z,\lambda,F\right)\leq0,\:\forall z\in\mathbb{R}_{++}.
$
\end{defn}

Given the stochastic dominance relation above, stochastic spanning
occurs when augmentation of the portfolio space does not enhance investment
opportunities, or equivalently, 
investment opportunities are not lost when the portfolio space is reduced. The following definition clarifies
the concept w.r.t.\ the Prospect dominance relation. 
\begin{defn}
\label{M-span}$\mathbb{K}$ Prospect-spans $\mathbb{L}$ ($\mathbb{K}\succcurlyeq_{P}\mathbb{L}$)
iff for any $\lambda\in\mathbb{L}$, $\exists\kappa\in\mathbb{K}:\kappa\succcurlyeq_{P}\lambda$.
If $\mathbb{K=\left\{ \kappa\right\} }$, the element $\kappa$ of
the singleton $\mathbb{K}$ is termed as Prospect super-efficient. 
\end{defn}

The efficient set of the dominance relation is the subset of $\mathbb{L}$
that contains the maximal elements. The efficient set is a spanning
subset of the portfolio space. Thereby, any superset of the efficient
set is also a spanning subset of $\mathbb{L}$. We can consider a
spanning set as an outer approximation of the efficient set. Given
a candidate spanning set exists, the question is whether this actually
spans the portfolio space. If a method for answering such a question
also exists, we can accurately approximate the efficient set via the
choice of finer spanning subsets of the portfolio space. This is important
in the context of decision theory and investment choice.

Hence, the question we address here is: given a candidate $\mathbb{\ensuremath{K}}$,
is $\mathbb{K}\succcurlyeq_{P}\mathbb{L}$? The following lemma provides
an analytical characterization by means of nested optimizations, which
is key for a numerical implementation on real data and statistical inference. 
\begin{lem}
\label{M-equiv-1}Suppose that $\mathbb{K}$ is closed. Then $\mathbb{K}\succcurlyeq_{P}\mathbb{L}$
iff we get the condition
$\displaystyle
\rho\left(F\right):=\max_{i=1,2}\sup_{\lambda\in\mathbb{L}}\sup_{z\in A_{i}}\inf_{\kappa\in\mathbb{K}}P_{i}\left(z,\lambda,\kappa,F\right)=0,
$
where $A_{1}=\mathbb{R}_{-},\:A_{2}=\mathbb{R}_{++}$. Moreover, we get that 
$\kappa$ is Prospect super-efficient iff $
\sup_{\lambda\in\mathbb{L}}\max_{i=1,2}\sup_{z\in A_{i}}P_{i}\left(z,\lambda,\kappa,F\right)=0.
$

\end{lem}

\subsection{Representation By Utility Functions}

We provide an expected utility characterization of spanning. Aside
the economic interpretation, this is key to the numerical LP implementation
of the inferential procedures that we construct in the next section. In doing so, we generalize
the utility characterization of PSD in Levy and Levy (2002), in
the sense that we do not require differentiability of the utilities. Our approach is in the spirit of the Russel and Seo (1989)
representations for the second order stochastic dominance. We rely on utilities
represented as unions of graphs of convex mixtures of appropriate
``ramp functions'' on each half-line.

To this end, we denote with $\mathcal{W}_{-},\mathcal{W}_{+}$, the
sets of Borel probability measures on the real line with supports
that are closed subsets of $\mathbb{R}_{-}$ and $\mathbb{R}_{+}$,
respectively, with existing first moments and uniformly integrable.
The latter requirement is convenient yet harmless
since orderings are invariant to utility rescalings. Those sets are
convex, and closed w.r.t.\ the topology of weak convergence and their
union contains the set of degenerate measures. Define 
$
V_{-}:=\left\{ v_{w}:\mathbb{R}_{-}\rightarrow\mathbb{R},\:v_{w}\left(u\right)=\int_{\mathbb{R}_{-}}\left[z1_{u\leq z}+u1_{z\leq u\leq0}\right]dw\left(z\right),\:w\in\mathcal{W}_{-}\right\} ,
$
and 
$
V_{+}:=\left\{ v_{w}:\mathbb{R}_{+}\rightarrow\mathbb{R},\:v_{w}\left(u\right)=\int_{\mathbb{R}_{+}}\left[u1_{0\leq u\leq z}+z1_{z\leq u<+\infty}\right]dw\left(z\right),\:w\in\mathcal{W}_{+}\right\} .
$
Every element of $V_{+}$ is increasing and concave, and dually every
element of $V_{-}$ is increasing and convex. Furthermore, any function
defined by the union of the graph of an arbitrary element of $V_{+}$
with the graph of an arbitrary element of $V_{-}$ is the graph of
an S-shaped utility function as defined by Levy and Levy (2002). Such
a utility function is concave for gains and convex for losses. Denote
the set of S-shaped utility functions obtained by such graph unions
as $V$. Thereby,

\[
V:=\left\{ v:\mathbb{R}\rightarrow\mathbb{R},\:v\left(u\right)=\begin{cases}
v_{w_{1}}\left(u\right), & u\leq0\\
v_{w_{2}}\left(u\right), & u\geq0
\end{cases}\mbox{\text{, where }}v_{w_{1}}\in V_{-},v_{w_{2}}\in V_{+}\right\} .
\]

\begin{lem}
\label{lem:ure}We have 
$
\rho\left(F\right)=\max_{i=1,2}\sup_{v_{w}\in V_{i}}\left[\sup_{\lambda\in\mathbb{L}}\mathbb{E}_{\lambda}\left[1_{u\in A_{i}}v_{w}\left(u\right)\right]-\sup_{\kappa\in\mathbb{K}}\mathbb{E}_{\kappa}\left[1_{u\in A_{i}}v_{w}\left(u\right)\right]\right],
$
where $\mathbb{E}_{\lambda}$ denotes expectation w.r.t. $G(z,\lambda,F)$.
 If the hypotheses of Lemma \ref{M-equiv-1} hold and $\mathbb{K}$
is convex, then $\mathbb{K}\succcurlyeq_{P}\mathbb{L}$ iff, 
$
\sup_{v\in V}\left[\sup_{\lambda\in\mathbb{L}}\mathbb{E}_{\lambda}\left[v\right]-\sup_{\kappa\in\mathbb{K}}\mathbb{E}_{\kappa}\left[v\right]\right]=0.
$
\end{lem}

The fist part of the lemma connects the functional that represents
spanning to the aforementioned classes of utilities. This is exploited
below in order to obtain feasible numerical formulations based on
LP. Those formulations are reminiscent of the LP programs developed
in the early papers of testing for SSD efficiency of a given portfolio
by Post (2003) and Kuosmanen (2004). The second part of Lemma \ref{lem:ure}
crystalizes the intuitive characterization of spanning w.r.t.\ investment
opportunities. It states that spanning holds if and only if the reduction
of investment opportunities from $\mathbb{L}$ to $\mathbb{K}$ does
not reduce optimal choices uniformly w.r.t.\ this class of preferences.

\subsection{An Asymptotically Exact and Consistent Test for Spanning}

We cannot directly rely on Lemma \ref{M-equiv-1} for empirical work if $F$ is unknown
and/or the optimizations are infeasible. We construct
a feasible statistical test for the null hypothesis of $\mathbb{K}\succcurlyeq_{P}\mathbb{L}$
by utilizing an empirical approximation of $F$ and by building feasible and fast
optimisations with LP. The null and alternative hypotheses take the
following forms: 
$
\mathbf{H_{0}}:\rho\left(F\right)=0,$ and $
\mathbf{H_{a}}:\rho\left(F\right)>0.$
In the special case of super-efficiency, the hypotheses
write as in Arvanitis and Topaloglou (2017).

We consider a process $\left(Y_{t}\right)_{t\in{\mathbb{Z}}}$ taking
values in $\mathbb{R}^{n}$. $Y_{i,t}$ denotes the $i^{th}$ element
of $Y_{t}$. The sample path of size $T$ is the random element $\left(Y_{t}\right)_{t=1,\ldots,T}$.
In our empirical finance framework, it represents returns of $n$
financial assets upon which we can construct portfolios via convex
combinations. $F$ is the cdf of $Y_{0}$ and $F_{T}$ is the empirical
cdf associated with the random element $\left(Y_{t}\right)_{t=1,\ldots,T}$.
Under our assumptions below, $F_{T}$ is a consistent
estimator of $F$, so we consider the following test statistic $
\rho_{T}:=\sqrt{T}\rho\left(F_{T}\right)=\sqrt{T}\max_{i=1,2}\sup_{\lambda\in\mathbb{L}}\sup_{z\in A_{i}}\inf_{\kappa\in\mathbb{K}}P_{i}\left(z,\lambda,\kappa,F_{T}\right),$
which is the scaled empirical analog of $\rho\left(F\right)$. As
already mentioned, when $\mathbb{K}$ is a singleton, the test statistic
coincides with the one used in Arvanitis and Topaloglou (2017). The
following assumption enables the derivation of the limit distribution
of $\rho_{T}$ under $\mathbf{H_{0}}$ and is weaker
than Assumption 2 in Arvanitis, Scaillet and Topaloglou (2019). 
\begin{assumption}
\label{MSDmix}$F$ is absolutely continuous w.r.t.\
the Lebesgue measure on $\mathbb{R}^{n}$ with convex support that
is bounded from below, and for some $0<\delta$, $\mathbb{E}\left[\left\Vert Y_{0}\right\Vert ^{2+\delta}\right]<+\infty$.
$\left(Y_{t}\right)_{t\in{\mathbb{Z}}}$ is $a$-mixing with mixing
coefficients $a_{T}=O(T^{-a})$ for some $a>1+\frac{2}{\eta},\:0<\eta<2$,
as $T\rightarrow\infty$.
\end{assumption}

The lower bound hypothesis is harmless in our empirical finance framework
since we are using financial returns. The mixing part is readily
implied by concepts such as geometric ergodicity which holds for many
stationary models used in the context of financial econometrics under
parameter restrictions and restrictions on the properties of the underlying
innovation processes. Examples are the strictly stationary versions
of (possibly multivariate) ARMA or several GARCH and stochastic volatility
type of models (see Francq and Zakoian (2011) for several examples).
Counter-examples are models that exhibit long memory, etc. The moment
condition is established in the aforementioned models via restrictions
on the properties of building blocks and the parameters of the processes
involved.

For the derivation of the limit theory of $\rho_{T}$
under the null hypothesis, we consider the contact sets 
$
\Gamma_{i}=\left\{ \lambda\in\mathbb{L},\kappa\in\mathbb{K}_{\lambda}^{\succeq},z\in A_{i}:P_{i}\left(z,\lambda,\kappa,F\right)=0\right\} ,
$
where $\mathbb{K}_{\lambda}^{\succeq}:=\left\{ \kappa\in\mathbb{K}:\kappa\succcurlyeq_{P}\lambda\right\} $
which under the null contains elements different from $\lambda$ for
any element of $\mathbb{L}-\mathbb{K}$. For any $i$, the set $\Gamma_{i}$
is non empty since $\Gamma_{i}^{\star}:=\left\{ \left(\kappa,\kappa,z\right),\kappa\in\mathbb{K},z\in A_{i}\right\} \subseteq\Gamma_{i}$.
Furthermore, $\left(\lambda,\kappa,0\right)\in\Gamma_{1},\:\forall\lambda,\kappa$.
Since due to Assumption \ref{MSDmix} $\underline{z}:=\inf_{\lambda,Y_{0}}\lambda'Y_{0}$
exists, for all $z\leq\underline{z}$, $\left(\lambda,\kappa,z\right)\in\Gamma_{i},\:\forall\lambda\in\mathbb{L},\kappa\in\mathbb{K}_{\lambda}^{\succeq}$
for the $i$ that corresponds to the sign of $\underline{z}$. 
In what follows, we denote convergence in distribution by $\rightsquigarrow$. 
\begin{prop}
\label{EAD}Suppose that $\mathbb{K}$ is closed, Assumption \ref{MSDmix}
holds and that \textup{$\mathbf{H_{0}}$} is true. Then as $T\rightarrow\infty$,
$
\rho_{T}\rightsquigarrow\rho_{\infty},
$
where 
$
\rho_{\infty}:=\max_{i=1,2}\sup_{\lambda}\sup_{z}\inf_{\kappa}P_{i}\left(z,\lambda,\kappa,\mathcal{G}_{F}\right),\:\left(\lambda,z,\kappa\right)\in\Gamma_{i},
$
and $\mathcal{G}_{F}$ is a centered Gaussian process with
covariance kernel given by \linebreak{}
$\text{Cov}(\mathcal{G}_{F}(x),\mathcal{G}_{F}(y))=\sum_{t\in\mathbb{Z}}\text{Cov}\left(1_{\{Y_{0}\leq x\}},1_{\{Y_{t}\leq y\}}\right)$
and $\mathbb{P}$ almost surely uniformly continuous sample paths
defined on $\mathbb{R}^{n}$.
\end{prop}

The limiting random variables have the form of saddle points of Gaussian
processes w.r.t.\ subsets of the relevant parameter spaces. This
is well defined since
$ \displaystyle
\mbox{\ensuremath{\text{Var}}}\int_{0}^{+\infty}\mathcal{G}_{\lambda F}\left(u\right)du=\int_{0}^{+\infty}\sum_{t\in\mathbb{Z}}\textnormal{Cov}\left(1_{\{\lambda^{T}Y_{0}\leq u\}},1_{\{\lambda^{Tr}Y_{t}\leq u\}}\right)du
$
$ \displaystyle
\leq2\sum_{t=0}^{\infty}\sqrt{a_{T}}\int_{0}^{+\infty}\sqrt{1-G\left(u,\lambda,F\right)}du<+\infty,
$
and 
$\displaystyle
\mbox{\ensuremath{\ensuremath{\text{Var}}}}\int_{-\infty}^{0}\mathcal{G}_{\lambda F}\left(u\right)du=\int_{-\infty}^{0}\sum_{t\in\mathbb{Z}}\textnormal{Cov}\left(1_{\{\lambda^{T}Y_{0}\leq u\}},1_{\{\lambda^{Tr}Y_{t}\leq u\}}\right)du
$
$\displaystyle
\leq2\sum_{t=0}^{\infty}\sqrt{a_{T}}\int_{-\infty}^{0}\sqrt{G\left(u,\lambda,F\right)}du$ $<+\infty,
$
where the first inequalities in each of the previous expressions follow
from inequality 1.12b in Rio (2000), and the second ones follow from
Assumption \ref{MSDmix} (see also p.\ 196 of Horvath et al.\ (2006)).\textcolor{red}{{} }

Since $F$ and $\Gamma_{i}$ are unknown in practice,
we use the results of the previous lemma to construct a decision procedure
based on subsampling, in the spirit of Linton, Post and Whang (2014) (see also Linton, Maasoumi, and Whang (2005)).\footnote{The partitioning used to get the results in Proposition \ref{EAD}
directly leads to the consideration of subsampling as a resampling
procedure. A testing procedure based on (block) bootstrap as in Scaillet
and Topaloglou (2010), can, due to the form of the recentering, be
consistent, but can be too conservative asymptotically, and thereby
suffer from a lack of power compared to the subsampling under particular
local alternatives (see also the relevant discussion in Arvanitis
et al.\ (2019)). The potential of asymptotic exactness for the subsampling
test justifies the particular resampling choice for inference.}
\begin{lyxalgorithm}
\label{Sub_alg}This consists of the following steps: 
\end{lyxalgorithm}

\begin{enumerate}
\item Evaluate $\rho_{T}$ at the original sample value. 
\item For $0<b_{T}\leq T$, generate subsample values \\ from the original observations
$(Y_{l})_{l=t,\ldots t+b_{T}-1}$ for all $t=1,2,\ldots,T-b_{T}+1$. 
\item Evaluate the test statistic on each subsample value \\ thereby obtaining
$\rho_{T,b_{T},t}$ for all $t=1,2,\ldots,T-b_{T}+1$. 
\item Approximate the cdf of the asymptotic distribution under the null
of $\rho_{T}$ \\ by $s_{T,b}(y)=\frac{1}{T-b_{T}+1}\sum_{t=1}^{T-b_{T}+1}1\left(\rho_{T,b_{T},t}\leq y\right)$
and calculate its $1-\alpha$ quantile \\
$
q_{T,b_{T}}\left(1-\alpha\right)=\inf_{y}\left\{ s_{T,b}(y)\geq1-\alpha\right\} ,
$ for the significance level $0< \alpha < .5$.
\item $\text{{Reject the null hypothesis}\:}\mathbf{H_{0}}\:\text{{if}\:}\rho_{T}>q_{T,b_{T}}(1-\alpha).$ 
\end{enumerate}
In order to derive the limit theory for the testing procedure, namely its asymptotic exactness and consistency stated in the next theorem, we first use
the following standard assumption that restricts the asymptotic behaviour
of  $b_{T}$ governing the size $b_{T}+1$ of each subsample. 
\begin{assumption}
\label{subseq}Suppose that $\left(b_{T}\right)$, possibly depending
on \textup{$\left(Y_{t}\right)_{t=1,\ldots,T}$, satisfies the condition  $\mathbb{P}\left(l_{T}\leq b_{T}\leq u_{T}\right)\rightarrow1$,
where $(l_{T})$ and $(u_{T})$ are real sequences such that $1\leq l_{T}\leq u_{T}$
for all $T$, $l_{T}\rightarrow\infty$ and $\frac{u_{T}}{T}\rightarrow0$
as $T\rightarrow\infty$.} 
\end{assumption}

\begin{thm}
\label{main2} Suppose Assumptions \ref{MSDmix} and \ref{subseq}
hold. For the testing procedure described in Algorithm \ref{Sub_alg}, we have
that 
\begin{enumerate}
\item If \textup{$\mathbf{H_{0}}$} is true, and for 
$\lambda\in\mathbb{L}-\mathbb{K}$, $\inf_{Y_{0}}\lambda^{Tr}Y_{0}\leq0$
there exists $\left(\kappa,z\right)\in\mathbb{K}_{\lambda}^{\succeq}\times\mathbb{R}_{++}$
with $\left(\lambda,\kappa,z\right)\in\Gamma_{2}$ and that if $\left(\lambda,\kappa^{\star},z^{\star}\right)\in\Gamma_{2}$
for  $\kappa^{\star}\neq\kappa$ then $z^{\star}\neq z$, then
for all $\alpha\in\left(0,.5\right)$ 
$
\lim_{T\rightarrow\infty}\mathbb{P}\left(\rho_{T}>q_{T,b_{T}}\left(1-\alpha\right)\right)=\alpha.
$
\item If \textup{$\mathbf{H_{a}}$} is true then 
$
\lim_{T\rightarrow\infty}\mathbb{P}\left(\rho_{T}>q_{T,b_{T}}\left(1-\alpha\right)\right)=1.
$
\end{enumerate}
\end{thm}

When for $\lambda\in\mathbb{L}-\mathbb{K}$,
$\inf_{Y_{0}}\lambda^{Tr}Y_{0}\leq0$ then due to Assumption \ref{MSDmix}
for any contact triple $\left(\lambda,\kappa,z\right)\in\Gamma_{2}$
we have that $P_{2}\left(z,\lambda,\kappa,\mathcal{G}_{F}\right)$
must be non-degenerate. Whenever $z$ corresponds solely to the particular
$\kappa$, we obtain that $\rho_{\infty}$ is non-degenerate and if
its cdf jumps at the infimum of its support, then the jump magnitude
is bounded above by $.5$. Hence in this case the test is
asymptotically exact for all the usual choices of the significance
level since  the probability of rejection under the null hypothesis, i.e., the size of the test, reaches $\alpha$ in large samples. 
We combine Proposition 6 above and Theorem 3.5.1
of Politis, Romano and Wolf (1999) in the proof of the exactness statement,  namely point 1 of Theorem \ref{main2}. 
To get exactness, the condition imposed on $\mathbb{L}-\mathbb{K}$  is significantly weaker
than the assumption on the relation between the extreme points of
$\mathbb{L}$ and $\mathbb{K}$ adopted by Arvanitis, Scaillet and
Topaloglou (2019).  It amounts to the existence of a spanned portfolio whose support is not strictly positive and so that, in the event of positive returns, there exists an elementary increasing and concave utility for positive returns and a unique portfolio such that the latter dominates the former and we are indifferent between the two portfolios with this particular utility.  Besides, the test is also consistent 
since the probability of rejection under the alternative hypothesis, i.e., the power of the test, reaches 1 in large samples.
We show in the proof of the consistency statement, namely point 2 of Theorem \ref{main2}, that the test statistic diverges to $+\infty$ under the alternative hypothesis
when $T$ goes to $+\infty$.

We opt for the ``bias correction'' regression analysis
of Arvanitis et al.\ (2019) to reduce the sensitivity of the quantile
estimates $q_{T,b_{T}}(1-\alpha)$ on the choice of $b_{T}$ in empirically
realistic dimensions for $n$ and $T$ (see also Arvanitis, Scaillet and
Topaloglou (2019) for further evidence on its better finite sample properties). Specifically, given $\alpha$,
we compute the quantiles $q_{T,b_{T}}(1-\alpha)$ for a  ``reasonable''
range of  $b_{T}$. Next, we estimate the intercept
and slope of the following regression line by OLS:
$
q_{T,b_{T}}(1-\alpha)=\gamma_{0;T,1-\alpha}+\gamma_{1;T,1-\alpha}(b_{T})^{-1}+\nu_{T;1-\alpha,b_{T}}.
$
Finally, we estimate the bias-corrected $(1-\alpha)$-quantile as
the OLS predicted value for $b_{T}=T$:
$
q_{T}^{BC}(1-\alpha):=\hat{\gamma}_{0;T,1-\alpha}+\hat{\gamma}_{1;T,1-\alpha}(T)^{-1}.\label{eq:estimated critical value}
$
Since $q_{T,b_{T}}(1-\alpha)$ converges in probability to $q(\rho_{\infty},1-\alpha)$
and $(b_{T})^{-1}$ converges to zero as $T\rightarrow0$, $\hat{\gamma}_{0;T,1-\alpha}$
converges in probability to $q(\rho_{\infty},1-\alpha)$ and the asymptotic
properties are not affected.

In the Online Appendix, we also show that under
further assumptions, the test is asymptotically locally unbiased under
given sequences of local alternatives. Besides, the Monte Carlo analysis reported in the Online Appendix shows that
the test performs well with an empirical size close to 5\%
and an empirical power above 90\% for a significance level $\alpha=5\%$.

\section{Numerical Implementation}

In this section, we exploit the results of Lemma \ref{lem:ure} in
order to provide with a finitary approximation of the test statistic.
We rely on this to provide with a numerical implementation based on
LP below.
We denote expectation w.r.t.\ the empirical measure by $\mathbb{E}_{F_{T}}$.
Let $\mathcal{R}^{-}$ denote
$\max_{i=1,\dots,n}\text{Range}\left(Y_{i,t}1_{Y_{i,t}\leq 0}\right)_{t=1,\dots,T}=\left[\underline{x},0\right]$.
Partition $\mathcal{R}^{-}$ into $n_{1}$ equally spaced values as
$\underline{x}=z_{1}<\cdots<z_{n_{1}}=0$, where $z_{n}:=\underline{x}-\frac{n-1}{n_{1}-1}\underline{x}$,
$n=1,\cdots,n_{1}$; $n_{1}\geq2$. Furthermore, partition the interval
$[0,1]$, as $0<\frac{1}{n_{2}-1}<\cdots<\frac{n_{2}-2}{n_{2}-1}<1$,
$n_{2}\geq2$. Similarly, $\mathcal{R}^{+}:=\max_{i=1,\dots,n}\text{Range}\left(Y_{i,t}1_{Y_{i,t}\geq 0}\right)_{t=1,\dots,T}=\left[0,\overline{x}\right]$.
Partition $\mathcal{R}^{+}$ into $p_{1}$ equally spaced values as
$0=z_{1}<\cdots<z_{p_{1}}=\overline{x}$, where $z_{p}:=\frac{p-1}{p_{1}-1}\overline{x}$,
$n=1,\cdots,p_{1}$; $p_{1}\geq2$, and again partition the interval
$[0,1]$, as $0<\frac{1}{p_{2}-1}<\cdots<\frac{p_{2}-2}{p_{2}-1}<1$,
$p_{2}\geq2$. Using the above, we consider the test statistic:
\begin{equation}\label{test_stat1}
\rho_{T}^{\star}:=\sqrt{T}\max_{i=1,2}\sup_{v\in V_{i}^{\star}}\left[\sup_{\lambda\in\mathbb{L}}\mathbb{E}_{F_{T}}\left[v\left(\lambda^{T}Y\right)\right]-\sup_{\kappa\in\mathbb{K}}\mathbb{E}_{F_{T}}\left[v\left(\kappa^{T}Y\right)\right]\right],
\end{equation}

where  the set of utility functions for negative returns is:
\[
V_{-}^{\star}:=\left\{ v:v(u)=\sum_{n=1}^{n_{1}}w_{n}\left[z_{n}1_{\underline{x}\leq u\leq z_{n}}+u1_{z_{n}\leq u\leq0}\right],\:\left(w_{1},\dots,w_{n_{1}}\right)\text{\ensuremath{\in}}\mathrm{W^{-}}\right\} ,
\]
\[
\mathrm{W^{-}:=\left\{ \left(w_{1},\dots,w_{n_{1}}\right)\in\left\{ 0,\frac{1}{n_{2}-1},\cdots,\frac{n_{2}-2}{n_{2}-1},1\right\} ^{n_{1}}:\sum_{n=1}^{n_{1}}w_{n}=1\right\} ,}
\]
and  the set of utility functions for positive returns  is:
\[
V_{+}^{\star}:=\left\{ v:v(u)=\sum_{p=1}^{p_{1}}w_{p}\left[u1_{0\leq u\leq z_{p}}+z_{p}1_{z_{p}\leq u\leq\overline{x}}\right],\:\left(w_{1},\dots,w_{p_{1}}\right)\text{\ensuremath{\in}}\mathrm{W^{+}}\right\} ,
\]
\[
\mathrm{W^{+}:=\left\{ \left(w_{1},\dots,w_{p_{1}}\right)\in\left\{ 0,\frac{1}{p_{2}-1},\cdots,\frac{p_{2}-2}{p_{2}-1},1\right\} ^{p_{1}}:\sum_{p=1}^{p_{1}}w_{p}=1\right\} .}
\]

We obtain the following result on the approximation of $\rho_{T}$
by $\rho_{T}^{\star}$. 
\begin{prop}
\label{prop:num1}When the support of $F$ is also
bounded from above, as $n_{1},n_{2},p_{1},p_{2}\rightarrow\infty$,
we have $\rho_{T}^{\star}\rightarrow\rho_{T}$, $\mathbb{P}$
a.s.
\end{prop}

Our feasible computational strategy builds on LP formulations for the numerical evaluation using the
previous finitary approximation of the test statistic.

We have a set of convex utility functions of the form:
$
v(u)=\sum_{n=1}^{n_{1}}w_{n} \max(u,z_n)
$ for the negative part.
For every $v \in V_{-}^{\star}$, we have at most $n_2$ line segments with knots at $n_1$ possible outcome levels. Then, we can enumerate all
$n_3=\frac{1}{(n_1-1)!} \prod_{i=1}^{n_1-1}(n_2+i-1)$ elements of $V_{-}^{\star}$. Our application in Section 4 uses $n_1=10$, and $n_2=5$, which gives $n_3=715$ distinct utility functions,
and a total of 1430 small LP problems for the two embedded maximisation problems in (\ref{test_stat1}). 
Solving (\ref{test_stat1}) yields simultaneously the optimal factor portfolio $\kappa$, and the optimal augmented portfolio $\lambda$ that maximize the expected utility.  Below, we give the mathematical formulation for the first optimization problem $\mathbb{\sup_{\lambda \in \mathrm{\Lambda}}}\mathbb{E}_{F_{N}}\left[u\left(\lambda^T Y \right)\right]$, that yields the optimal augmented portfolio $\lambda$.
The same formulation is used for the second optimization $\mathbb{\sup_{\kappa \in \mathrm{\kappa}}}\mathbb{E}_{F_{N}}\left[u\left(\kappa^T Y \right)\right]$.

Let us  define:
$c_{0,n}:=\sum_{m=n}^{n_1}\left(c_{1,m}-c_{1,m+1}\right)z_{m}$, 
$c_{1,n}:=\sum_{m=n}^{n_1}w_{m}$, and \break
$\mathcal{N}:=\left \{ n=1,\cdots,n_{1}:w_{n}>0\right \} \bigcup \left \{ n_{1}\right \} $.
For any given $u\in V_{-}$,  $\mathbb{\sup_{\lambda \in \mathrm{\Lambda}}}\mathbb{E}_{F_{N}}\left[u\left(\lambda^T Y \right)\right]$
is the optimal value of the objective function of the following LP
problem in canonical form:
\begin{gather}
\max T^{-1}\sum_{t=1}^{T}y_{t}\label{eq:Opt1}\\
\text{s.t., for }\nonumber t=1,\cdots, T, \quad n\in \mathcal{N}, \quad i=1,\cdots,M, \\
y_{t} \leq \lambda^T Y_t c_{1,n}   + Q_t^{-} + Q_t^{+}, \qquad 
y_{t} \leq c_{0,n}   + Q_t^{-} + Q_t^{+},\nonumber \\
Q_{t}^{-} \geq c_{0,n}   -\lambda^T Y_t c_{1,n}, \qquad
Q_{t}^{+} \geq \lambda^T Y_t c_{1,n} - c_{0,n}, \qquad Q_{t}^{-} \geq0, \qquad  Q_{t}^{+}, \geq0,\nonumber \\ 
\sum_{i=1}^{M}\lambda_{i}=1, \qquad \lambda_{i}\geq 0, \qquad \text{and }\quad  y_{t}\mathrm{\text{ being free}}.\nonumber
\end{gather}

We have a set of concave utility functions of the form:
$
v(u)=\sum_{p=1}^{p_{1}}w_{p} \min(u,z_p),
$ for the positive part.
Again, for every $v \in V_{+}^{\star}$, we have at most $p_2$ line segments with knots at $p_1$ possible outcome levels. As before, the number of elements of $V_{+}^{\star}$ is
$p_3=\frac{1}{(p_1-1)!} \prod_{i=1}^{p_1-1}(p_2+i-1)=1430$, for $p_1=10$ and $p_2=5$.

Let us define:
$
c_{0,p}:=\sum_{m=p}^{p_1}\left(c_{1,m}-c_{1,m+1}\right)z_{m}$, $c_{1,p}:=\sum_{m=p}^{p_1}w_{m}$, and \break $
\mathcal{P}:=\left \{ p=1,\cdots,p_{1}:w_{p}>0\right \} \bigcup \left \{ p_{1}\right \}$.
For any given $u\in V_{+}$,
$\mathbb{\sup_{\lambda \in \mathrm{\Lambda}}}\mathbb{E}_{F_{N}}\left[u\left(\lambda^T Y \right)\right]$
is the optimal value of the objective function of the following LP
problem in canonical form:
\begin{gather}
\max T^{-1}\sum_{t=1}^{T}y_{t}\label{eq:Opt2}\\
\text{s.t., for }\nonumber t=1,\cdots, T, \quad n\in \mathcal{P}, \quad i=1,\cdots,M, \\
y_{t} \leq \lambda^T Y_t c_{1,p}, \qquad 
y_{t} \leq c_{0,p},\qquad 
\sum_{i=1}^{M}\lambda_{i}=1\qquad \lambda_{i}\geq0,
 \qquad \text{and }\quad  y_{t}\mathrm{\text{ being free}}. \nonumber
\end{gather}

 The total run time for each computation does not exceed one minute when we use a desktop PC with a 3.6 GHz, 6-core Intel i7 processor, with 16 GB of RAM, using MATLAB and GAMS with the Gurobi optimization solver. 

\section{Empirical Application}

In the empirical application, we examine if we can explain well-known stock market
anomalies by standard factors within a new breed
of asset pricing models, for prospect type investor preferences. For
this purpose, we use the prospect spanning tests, both in- and out-of-sample.

\subsection{Factor Models and Anomalies}

We start with a benchmark factor model from a set of models that have generated support in the recent
literature, and we ask whether a characteristic identified in the literature as stock market anomaly,
is a market anomaly for prospect investors.  
To answer this question, we consider three models that build on the pioneer three-factor model
of Fama and French (1993): the four-factor model of Hou, Xue and Zhang
(2015), the five-factor model of Fama and French (2015), and the four-factor
model of Stambaugh and Yuan (2017). Fama and French (1993) aim to
capture the part of average stock returns left unexplained in CAPM
of Sharpe (1964) and Lintner (1965) by including, in addition to the market
factor, two extra risk factors relating to size (measured by market
equity) and the ratio of book-to-market equity. In addition to the market excess return,
the influential three-factor model of Fama and French (1993) includes a book-to-market
or "value" factor, HML, and a size factor, SMB, based on market capitalization. Motivated by Miller
and Modigliani (1961), Fama and French (2015) five-factor model (henceforth,
FF-5) augments the original Fama-French three-factor model by two
extra factors, one for profitability and another for investment. Hou,
Xue and Zhang (2015) consider a four-factor model (dubbed the q-factor
model) that includes the original market and size factors of Fama
and French (1993) augmented by a profitability and investment factor.
Stambaugh and Yuan (2017) consider a four-factor model (henceforth,
M-4) including the standard market and size factors along with two
composite factors for investment and profitability. To construct the
composite factors, they combine information from 11 market anomalies
relating to investment and profitability measures. We
use alternative factor models as a robustness check, namely for testing
the consistency of in- and out-of-sample results under the prospect
preferences, and not for a horse race in cross-sectional asset pricing.

The stock market anomalies we examine in this paper have a long history
in the relevant literature. A common theme in the original papers
that first highlighted these patterns, is that they all challenge
the rational asset pricing paradigm as they exhibit returns that are
not in line with the risks taken. However, notwithstanding whether
they are caused by sentiment (a catch-all term that stand for all
kinds of irrational decision-making) or by market frictions (e.g.
margin requirements), it is also acknowledged that most of them persist
because they cannot be ``arbitraged'' away. From the perspective
of the Arbitrage Pricing Theory this implies that arbitrageurs cannot
trade against them without exposing themselves to significant risks.
In this paper, we test the 11 strategies used to construct Stambaugh-Yuan
factors, along with Betting against Beta, Quality minus Junk, Size,
Growth Option, Value (Book to Market), Idiosyncratic volatility and
Profitability. The 11 anomalies used in Stambaugh and Yuan (2017)
are  Accruals, Asset Growth, Composite Equity Issue, Distress, Growth Profitability Premium,
Investment to Assets, Momentum, Net Operating Assets, Net Stock Issues, O-Score, and Return on Assets.
They are realigned appropriately to yield positive average returns. In
particular, anomaly variables that relate to investment activity (Asset
Growth, Investment to Assets, Net Stock Issues, Composite Equity Isues,
Aaccruals) are defined low-minus-high decile portfolio returns,
rather than high-minus-low, as in Hou et al.\ (2015). All the other
anomalies are constructed as high-minus-low decile portfolio returns.
A short description of the 18 market anomalies that we study in the paper
is given in Appendix A (see Stambaugh and Yuan (2017) for further details). Returns of the Fama and French 5 factors were downloaded from Kenneth
French's site. The dataset consists of all monthly observations
from January 1974 until December 2016. M-4 factor returns and anomaly
spread return series were downloaded from the websites of Robert Stambaugh
and AQR.
In the Online Appendix, we report summary statistics of the factor and anomaly returns over our sample period.

\subsection{In-Sample Analysis}

In this section, we test in-sample the null hypothesis that the set
of standard factors prospect spans the set enlarged with a particular
market anomaly. We test separately for the Fama and French 5 factors,
the Stambaugh-Yuan 4 factors as well as Hou-Xue-Zhang 4 factors, with
respect to each one of the 18 additional anomalies. We get the subsampling
distribution of the test statistic for subsample size $b_{T}\in \{T^{0.6},T^{0.7},T^{0.8},T^{0.9}\}$.
Using OLS regression on the empirical quantiles $q_{T,b_{T}}(1-\alpha)$
 for a significance level $\alpha=5\%$, we get the estimate $q_{T}^{BC}$
for the bias-corrected critical value. We reject spanning if the test statistic $\rho_{T}^{\star}$
is higher than the regression estimate $q_{T}^{BC}$.

Tables \ref{t2}-\ref{t4} report the test statistics $\rho_{T}^{\star}$
as well as the regression estimates $q_{T}^{BC}$ when we test for
spanning of the alternative factor models w.r.t.\ each one
of the 18 market anomalies.

\begin{table}
\caption{Test statistics: Fama and French (FF-5) Factors}
\label{t2}{\small{}{}{}{}{}{}{}{}{}{}{}{}{}{}{}{}{}\centering
}%
\begin{tabular}{lccc}
\hline 
{\small{}{}{}{}{}{}{}{}{}{}{}{}{}{}{}{}{}Variable}  & {\small{}{}{}{}{}{}{}{}{}{}{}{}{}{}{}{}{}Test statistic
$\rho_{T}^{\star}$}  & {\small{}{}{}{}{}{}{}{}{}{}{}{}{}{}{}{}{} Regression
estimates $q_{T}^{BC}$}  & {\small{}{}{}{}{}{}{}{}{}{}{}{}{}{}{}{}{} Result}\tabularnewline
\hline 
{\small{}{}{}{}{}{}{}{}{}{}{}{}{}{}{}{}{}Accruals}  & {\small{}{}{}{}{}{}{}{}{}{}{}{}{}{}{}{}{} 0.0016}  & {\small{}{}{}{}{}{}{}{}{}{}{}{}{}{}{}{}{} 0.0025}  & {\small{}{}{}{}{}{}{}{}{}{}{}{}{}{}{}{}{} Spanning}\tabularnewline
{\small{}{}{}{}{}{}{}{}{}{}{}{}{}{}{}{}{}Asset Growth}  & {\small{}{}{}{}{}{}{}{}{}{}{}{}{}{}{}{}{} 0.0}  & {\small{}{}{}{}{}{}{}{}{}{}{}{}{}{}{}{}{} 0.0}  & {\small{}{}{}{}{}{}{}{}{}{}{}{}{}{}{}{}{} Spanning}\tabularnewline
{\small{}{}{}{}{}{}{}{}{}{}{}{}{}{}{}{}{}Composite Equity
Issue}  & {\small{}{}{}{}{}{}{}{}{}{}{}{}{}{}{}{}{} 0.0015}  & {\small{}{}{}{}{}{}{}{}{}{}{}{}{}{}{}{}{} 0.0003}  & {\small{}{}{}{}{}{}{}{}{}{}{}{}{}{}{}{}{} Reject Spanning}\tabularnewline
{\small{}{}{}{}{}{}{}{}{}{}{}{}{}{}{}{}{}Distress}  & {\small{}{}{}{}{}{}{}{}{}{}{}{}{}{}{}{}{} 0.0045}  & {\small{}{}{}{}{}{}{}{}{}{}{}{}{}{}{}{}{} 0.0005}  & {\small{}{}{}{}{}{}{}{}{}{}{}{}{}{}{}{}{} Reject Spanning}\tabularnewline
{\small{}{}{}{}{}{}{}{}{}{}{}{}{}{}{}{}{}Growth Profitability Premium}  & {\small{}{}{}{}{}{}{}{}{}{}{}{}{}{}{}{}{} 0.0015}  & {\small{}{}{}{}{}{}{}{}{}{}{}{}{}{}{}{}{} 0.0012}  & {\small{}{}{}{}{}{}{}{}{}{}{}{}{}{}{}{}{} Reject Spanning}\tabularnewline
{\small{}{}{}{}{}{}{}{}{}{}{}{}{}{}{}{}{}Investment to Assets}  & {\small{}{}{}{}{}{}{}{}{}{}{}{}{}{}{}{}{} 0.0014}  & {\small{}{}{}{}{}{}{}{}{}{}{}{}{}{}{}{}{} 0.0001}  & {\small{}{}{}{}{}{}{}{}{}{}{}{}{}{}{}{}{} Reject Spanning}\tabularnewline
{\small{}{}{}{}{}{}{}{}{}{}{}{}{}{}{}{}{}Momentum}  & {\small{}{}{}{}{}{}{}{}{}{}{}{}{}{}{}{}{} 0.0696}  & {\small{}{}{}{}{}{}{}{}{}{}{}{}{}{}{}{}{} 0.0204}  & {\small{}{}{}{}{}{}{}{}{}{}{}{}{}{}{}{}{} Reject Spanning}\tabularnewline
{\small{}{}{}{}{}{}{}{}{}{}{}{}{}{}{}{}{}Net Operating Assets}  & {\small{}{}{}{}{}{}{}{}{}{}{}{}{}{}{}{}{} 0.0268}  & {\small{}{}{}{}{}{}{}{}{}{}{}{}{}{}{}{}{} 0.0009}  & {\small{}{}{}{}{}{}{}{}{}{}{}{}{}{}{}{}{} Reject Spanning}\tabularnewline
{\small{}{}{}{}{}{}{}{}{}{}{}{}{}{}{}{}{}Net Stock Issues}  & {\small{}{}{}{}{}{}{}{}{}{}{}{}{}{}{}{}{} 0.0011}  & {\small{}{}{}{}{}{}{}{}{}{}{}{}{}{}{}{}{} 0.0003}  & {\small{}{}{}{}{}{}{}{}{}{}{}{}{}{}{}{}{} Reject Spanning}\tabularnewline
{\small{}{}{}{}{}{}{}{}{}{}{}{}{}{}{}{}{}O-Score}  & {\small{}{}{}{}{}{}{}{}{}{}{}{}{}{}{}{}{} 0.0129}  & {\small{}{}{}{}{}{}{}{}{}{}{}{}{}{}{}{}{} 0.0092}  & {\small{}{}{}{}{}{}{}{}{}{}{}{}{}{}{}{}{} Reject Spanning}\tabularnewline
{\small{}{}{}{}{}{}{}{}{}{}{}{}{}{}{}{}{}Return on Assets}  & {\small{}{}{}{}{}{}{}{}{}{}{}{}{}{}{}{}{} 0.0024}  & {\small{}{}{}{}{}{}{}{}{}{}{}{}{}{}{}{}{} 0.0047}  & {\small{}{}{}{}{}{}{}{}{}{}{}{}{}{}{}{}{} Spanning}\tabularnewline
{\small{}{}{}{}{}{}{}{}{}{}{}{}{}{}{}{}{}Betting against Beta}  & {\small{}{}{}{}{}{}{}{}{}{}{}{}{}{}{}{}{} 0.0235}  & {\small{}{}{}{}{}{}{}{}{}{}{}{}{}{}{}{}{} 0.0176}  & {\small{}{}{}{}{}{}{}{}{}{}{}{}{}{}{}{}{} Reject Spanning}\tabularnewline
{\small{}{}{}{}{}{}{}{}{}{}{}{}{}{}{}{}{}Quality minus Junk}  & {\small{}{}{}{}{}{}{}{}{}{}{}{}{}{}{}{}{} 0.0088}  & {\small{}{}{}{}{}{}{}{}{}{}{}{}{}{}{}{}{} 0.0061}  & {\small{}{}{}{}{}{}{}{}{}{}{}{}{}{}{}{}{} Reject Spanning}\tabularnewline
{\small{}{}{}{}{}{}{}{}{}{}{}{}{}{}{}{}{}Size}  & {\small{}{}{}{}{}{}{}{}{}{}{}{}{}{}{}{}{} 0.0}  & {\small{}{}{}{}{}{}{}{}{}{}{}{}{}{}{}{}{} 0.0}  & {\small{}{}{}{}{}{}{}{}{}{}{}{}{}{}{}{}{} Spanning}\tabularnewline
{\small{}{}{}{}{}{}{}{}{}{}{}{}{}{}{}{}{}Growth Option}  & {\small{}{}{}{}{}{}{}{}{}{}{}{}{}{}{}{}{} 0.0}  & {\small{}{}{}{}{}{}{}{}{}{}{}{}{}{}{}{}{} 0.0}  & {\small{}{}{}{}{}{}{}{}{}{}{}{}{}{}{}{}{} Spanning}\tabularnewline
{\small{}{}{}{}{}{}{}{}{}{}{}{}{}{}{}{}{}Value (Book
to Market)}  & {\small{}{}{}{}{}{}{}{}{}{}{}{}{}{}{}{}{} 0.1921}  & {\small{}{}{}{}{}{}{}{}{}{}{}{}{}{}{}{}{} 0.1878}  & {\small{}{}{}{}{}{}{}{}{}{}{}{}{}{}{}{}{} Reject Spanning}\tabularnewline
{\small{}{}{}{}{}{}{}{}{}{}{}{}{}{}{}{}{}Idiosyncratic
Volatility}  & {\small{}{}{}{}{}{}{}{}{}{}{}{}{}{}{}{}{} 01959}  & {\small{}{}{}{}{}{}{}{}{}{}{}{}{}{}{}{}{} 0.0100}  & {\small{}{}{}{}{}{}{}{}{}{}{}{}{}{}{}{}{} Reject Spanning}\tabularnewline
{\small{}{}{}{}{}{}{}{}{}{}{}{}{}{}{}{}{}Profitability}  & {\small{}{}{}{}{}{}{}{}{}{}{}{}{}{}{}{}{} 0.0}  & {\small{}{}{}{}{}{}{}{}{}{}{}{}{}{}{}{}{} 0.0}  & {\small{}{}{}{}{}{}{}{}{}{}{}{}{}{}{}{}{} Spanning}\tabularnewline
\hline 
\end{tabular} 
\noindent %
\noindent\parbox[t][0.7\totalheight]{1\textwidth}{%
Entries report the 
test statistics $\rho_{T}^{\star}$
and the regression estimates $q_{T}^{BC}$  for
spanning of the Fama and French (FF-5) model with respect to each one
of the 18 market anomalies.
We reject spanning at significance level $\alpha=5\%$ if $\rho_{T}^{\star} > q_{T}^{BC}$.
The dataset spans the period
from January, 1974 to December, 2016.
} 
\end{table}

\begin{table}[H]
\caption{Test statistics: Stambaugh-Yuan (M-4) Factors}
\label{t3}{\small{}{}{}{}{}{}{}{}{}{}{}{}{}{}{}{}{}\centering
}%
\begin{tabular}{lccc}
\hline 
{\small{}{}{}{}{}{}{}{}{}{}{}{}{}{}{}{}{}Variable}  & {\small{}{}{}{}{}{}{}{}{}{}{}{}{}{}{}{}{}Test statistic
$\rho_{T}^{\star}$}  & {\small{}{}{}{}{}{}{}{}{}{}{}{}{}{}{}{}{} Regression
estimates $q_{T}^{BC}$}  & {\small{}{}{}{}{}{}{}{}{}{}{}{}{}{}{}{}{} Result}\tabularnewline
\hline 
{\small{}{}{}{}{}{}{}{}{}{}{}{}{}{}{}{}{}Accruals}  & {\small{}{}{}{}{}{}{}{}{}{}{}{}{}{}{}{}{} 0.0081}  & {\small{}{}{}{}{}{}{}{}{}{}{}{}{}{}{}{}{} 0.0083}  & {\small{}{}{}{}{}{}{}{}{}{}{}{}{}{}{}{}{} Spanning}\tabularnewline
{\small{}{}{}{}{}{}{}{}{}{}{}{}{}{}{}{}{}Asset Growth}  & {\small{}{}{}{}{}{}{}{}{}{}{}{}{}{}{}{}{} 0.0057}  & {\small{}{}{}{}{}{}{}{}{}{}{}{}{}{}{}{}{} 0.0069}  & {\small{}{}{}{}{}{}{}{}{}{}{}{}{}{}{}{}{} Spanning}\tabularnewline
{\small{}{}{}{}{}{}{}{}{}{}{}{}{}{}{}{}{}Composite Equity
Issue}  & {\small{}{}{}{}{}{}{}{}{}{}{}{}{}{}{}{}{} 0.0143}  & {\small{}{}{}{}{}{}{}{}{}{}{}{}{}{}{}{}{} 0.078}  & {\small{}{}{}{}{}{}{}{}{}{}{}{}{}{}{}{}{} Reject Spanning}\tabularnewline
{\small{}{}{}{}{}{}{}{}{}{}{}{}{}{}{}{}{}Distress}  & {\small{}{}{}{}{}{}{}{}{}{}{}{}{}{}{}{}{} 0.0533}  & {\small{}{}{}{}{}{}{}{}{}{}{}{}{}{}{}{}{} 0.0020}  & {\small{}{}{}{}{}{}{}{}{}{}{}{}{}{}{}{}{} Reject Spanning}\tabularnewline
{\small{}{}{}{}{}{}{}{}{}{}{}{}{}{}{}{}{}Growth Profitability Premium}  & {\small{}{}{}{}{}{}{}{}{}{}{}{}{}{}{}{}{} 0.0113}  & {\small{}{}{}{}{}{}{}{}{}{}{}{}{}{}{}{}{} 0.0049}  & {\small{}{}{}{}{}{}{}{}{}{}{}{}{}{}{}{}{} Reject Spanning}\tabularnewline
{\small{}{}{}{}{}{}{}{}{}{}{}{}{}{}{}{}{}Investment to Assets}  & {\small{}{}{}{}{}{}{}{}{}{}{}{}{}{}{}{}{} 0.0116}  & {\small{}{}{}{}{}{}{}{}{}{}{}{}{}{}{}{}{} 0.0164}  & {\small{}{}{}{}{}{}{}{}{}{}{}{}{}{}{}{}{} Reject Spanning}\tabularnewline
{\small{}{}{}{}{}{}{}{}{}{}{}{}{}{}{}{}{}Momentum}  & {\small{}{}{}{}{}{}{}{}{}{}{}{}{}{}{}{}{} 0.1189}  & {\small{}{}{}{}{}{}{}{}{}{}{}{}{}{}{}{}{} 0.1143}  & {\small{}{}{}{}{}{}{}{}{}{}{}{}{}{}{}{}{} Reject Spanning}\tabularnewline
{\small{}{}{}{}{}{}{}{}{}{}{}{}{}{}{}{}{}Net Operating Assets}  & {\small{}{}{}{}{}{}{}{}{}{}{}{}{}{}{}{}{} 0.0653}  & {\small{}{}{}{}{}{}{}{}{}{}{}{}{}{}{}{}{} 0.0071}  & {\small{}{}{}{}{}{}{}{}{}{}{}{}{}{}{}{}{} Reject Spanning}\tabularnewline
{\small{}{}{}{}{}{}{}{}{}{}{}{}{}{}{}{}{}Net Stock Issues}  & {\small{}{}{}{}{}{}{}{}{}{}{}{}{}{}{}{}{} 0.0145}  & {\small{}{}{}{}{}{}{}{}{}{}{}{}{}{}{}{}{} 0.0073}  & {\small{}{}{}{}{}{}{}{}{}{}{}{}{}{}{}{}{} Reject Spanning}\tabularnewline
{\small{}{}{}{}{}{}{}{}{}{}{}{}{}{}{}{}{}O-Score}  & {\small{}{}{}{}{}{}{}{}{}{}{}{}{}{}{}{}{} 0.0133}  & {\small{}{}{}{}{}{}{}{}{}{}{}{}{}{}{}{}{} 0.0122}  & {\small{}{}{}{}{}{}{}{}{}{}{}{}{}{}{}{}{} Reject Spanning}\tabularnewline
{\small{}{}{}{}{}{}{}{}{}{}{}{}{}{}{}{}{}Return on Assets}  & {\small{}{}{}{}{}{}{}{}{}{}{}{}{}{}{}{}{} 0.0012}  & {\small{}{}{}{}{}{}{}{}{}{}{}{}{}{}{}{}{} 0.0015}  & {\small{}{}{}{}{}{}{}{}{}{}{}{}{}{}{}{}{} Spanning}\tabularnewline
{\small{}{}{}{}{}{}{}{}{}{}{}{}{}{}{}{}{}Betting against Beta}  & {\small{}{}{}{}{}{}{}{}{}{}{}{}{}{}{}{}{} 0.0755}  & {\small{}{}{}{}{}{}{}{}{}{}{}{}{}{}{}{}{} 0.0703}  & {\small{}{}{}{}{}{}{}{}{}{}{}{}{}{}{}{}{} Reject Spanning}\tabularnewline
{\small{}{}{}{}{}{}{}{}{}{}{}{}{}{}{}{}{}Quality minus Junk}  & {\small{}{}{}{}{}{}{}{}{}{}{}{}{}{}{}{}{} 0.0374}  & {\small{}{}{}{}{}{}{}{}{}{}{}{}{}{}{}{}{} 0.0099}  & {\small{}{}{}{}{}{}{}{}{}{}{}{}{}{}{}{}{} Reject Spanning}\tabularnewline
{\small{}{}{}{}{}{}{}{}{}{}{}{}{}{}{}{}{}Size}  & {\small{}{}{}{}{}{}{}{}{}{}{}{}{}{}{}{}{} 0.0}  & {\small{}{}{}{}{}{}{}{}{}{}{}{}{}{}{}{}{} 0.0}  & {\small{}{}{}{}{}{}{}{}{}{}{}{}{}{}{}{}{} Spanning}\tabularnewline
{\small{}{}{}{}{}{}{}{}{}{}{}{}{}{}{}{}{}Growth Option}  & {\small{}{}{}{}{}{}{}{}{}{}{}{}{}{}{}{}{} 0.0}  & {\small{}{}{}{}{}{}{}{}{}{}{}{}{}{}{}{}{} 0.0}  & {\small{}{}{}{}{}{}{}{}{}{}{}{}{}{}{}{}{} Spanning}\tabularnewline
{\small{}{}{}{}{}{}{}{}{}{}{}{}{}{}{}{}{}Value (Book
to Market)}  & {\small{}{}{}{}{}{}{}{}{}{}{}{}{}{}{}{}{} 0.2939}  & {\small{}{}{}{}{}{}{}{}{}{}{}{}{}{}{}{}{} 0.2817}  & {\small{}{}{}{}{}{}{}{}{}{}{}{}{}{}{}{}{} Reject Spanning}\tabularnewline
{\small{}{}{}{}{}{}{}{}{}{}{}{}{}{}{}{}{}Idiosyncratic
Volatility}  & {\small{}{}{}{}{}{}{}{}{}{}{}{}{}{}{}{}{} 0.2593}  & {\small{}{}{}{}{}{}{}{}{}{}{}{}{}{}{}{}{} 0.1039}  & {\small{}{}{}{}{}{}{}{}{}{}{}{}{}{}{}{}{} Reject Spanning}\tabularnewline
{\small{}{}{}{}{}{}{}{}{}{}{}{}{}{}{}{}{}Profitability}  & {\small{}{}{}{}{}{}{}{}{}{}{}{}{}{}{}{}{} 0.0}  & {\small{}{}{}{}{}{}{}{}{}{}{}{}{}{}{}{}{} 0.0}  & {\small{}{}{}{}{}{}{}{}{}{}{}{}{}{}{}{}{} Spanning}\tabularnewline
\hline 
\end{tabular} 
\noindent %
\noindent\parbox[t][0.7\totalheight]{1\textwidth}{%
Entries report the 
test statistics $\rho_{T}^{\star}$
and the regression estimates $q_{T}^{BC}$  for
spanning of the Stambaugh-Yuan (M-4) model with respect to each one
of the 18 market anomalies.
We reject spanning at significance level $\alpha=5\%$ if $\rho_{T}^{\star} > q_{T}^{BC}$.
The dataset spans the period
from January, 1974 to December, 2016.
} 
\end{table}

\begin{table}[H]
\caption{Test statistics: Hou-Xue-Zhang (q) Factors}
\label{t4}{\small{}{}{}{}{}{}{}{}{}{}{}{}{}{}{}{}{}\centering
}%
\begin{tabular}{lccc}
\hline 
{\small{}{}{}{}{}{}{}{}{}{}{}{}{}{}{}{}{}Variable}  & {\small{}{}{}{}{}{}{}{}{}{}{}{}{}{}{}{}{}Test statistic
$\rho_{T}^{\star}$}  & {\small{}{}{}{}{}{}{}{}{}{}{}{}{}{}{}{}{} Regression
estimates $q_{T}^{BC}$}  & {\small{}{}{}{}{}{}{}{}{}{}{}{}{}{}{}{}{} Result}\tabularnewline
\hline 
{\small{}{}{}{}{}{}{}{}{}{}{}{}{}{}{}{}{}Accruals}  & {\small{}{}{}{}{}{}{}{}{}{}{}{}{}{}{}{}{} 0.0106}  & {\small{}{}{}{}{}{}{}{}{}{}{}{}{}{}{}{}{} 0.0039}  & {\small{}{}{}{}{}{}{}{}{}{}{}{}{}{}{}{}{} Reject Spannin}\tabularnewline
{\small{}{}{}{}{}{}{}{}{}{}{}{}{}{}{}{}{}Asset Growth}  & {\small{}{}{}{}{}{}{}{}{}{}{}{}{}{}{}{}{} 0.0176}  & {\small{}{}{}{}{}{}{}{}{}{}{}{}{}{}{}{}{} 0.0101}  & {\small{}{}{}{}{}{}{}{}{}{}{}{}{}{}{}{}{} Reject Spanning}\tabularnewline
{\small{}{}{}{}{}{}{}{}{}{}{}{}{}{}{}{}{}Composite Equity
Issue}  & {\small{}{}{}{}{}{}{}{}{}{}{}{}{}{}{}{}{} 0.0163}  & {\small{}{}{}{}{}{}{}{}{}{}{}{}{}{}{}{}{} 0.0159}  & {\small{}{}{}{}{}{}{}{}{}{}{}{}{}{}{}{}{} Reject Spanning}\tabularnewline
{\small{}{}{}{}{}{}{}{}{}{}{}{}{}{}{}{}{}Distress}  & {\small{}{}{}{}{}{}{}{}{}{}{}{}{}{}{}{}{} 0.0386}  & {\small{}{}{}{}{}{}{}{}{}{}{}{}{}{}{}{}{} 0.0133}  & {\small{}{}{}{}{}{}{}{}{}{}{}{}{}{}{}{}{} Reject Spanning}\tabularnewline
{\small{}{}{}{}{}{}{}{}{}{}{}{}{}{}{}{}{}Growth Profitability Premium}  & {\small{}{}{}{}{}{}{}{}{}{}{}{}{}{}{}{}{} 0.0084}  & {\small{}{}{}{}{}{}{}{}{}{}{}{}{}{}{}{}{} 0.0038}  & {\small{}{}{}{}{}{}{}{}{}{}{}{}{}{}{}{}{} Reject Spanning}\tabularnewline
{\small{}{}{}{}{}{}{}{}{}{}{}{}{}{}{}{}{}Investment to Assets}  & {\small{}{}{}{}{}{}{}{}{}{}{}{}{}{}{}{}{} 0.0157}  & {\small{}{}{}{}{}{}{}{}{}{}{}{}{}{}{}{}{} 0.0123}  & {\small{}{}{}{}{}{}{}{}{}{}{}{}{}{}{}{}{} Reject Spanning}\tabularnewline
{\small{}{}{}{}{}{}{}{}{}{}{}{}{}{}{}{}{}Momentum}  & {\small{}{}{}{}{}{}{}{}{}{}{}{}{}{}{}{}{} 0.0835}  & {\small{}{}{}{}{}{}{}{}{}{}{}{}{}{}{}{}{} 0.0305}  & {\small{}{}{}{}{}{}{}{}{}{}{}{}{}{}{}{}{} Reject Spanning}\tabularnewline
{\small{}{}{}{}{}{}{}{}{}{}{}{}{}{}{}{}{}Net Operating Assets}  & {\small{}{}{}{}{}{}{}{}{}{}{}{}{}{}{}{}{} 0.0449}  & {\small{}{}{}{}{}{}{}{}{}{}{}{}{}{}{}{}{} 0.0059}  & {\small{}{}{}{}{}{}{}{}{}{}{}{}{}{}{}{}{} Reject Spanning}\tabularnewline
{\small{}{}{}{}{}{}{}{}{}{}{}{}{}{}{}{}{}Net Stock Issues}  & {\small{}{}{}{}{}{}{}{}{}{}{}{}{}{}{}{}{} 0.0178}  & {\small{}{}{}{}{}{}{}{}{}{}{}{}{}{}{}{}{} 0.0170}  & {\small{}{}{}{}{}{}{}{}{}{}{}{}{}{}{}{}{} Reject Spanning}\tabularnewline
{\small{}{}{}{}{}{}{}{}{}{}{}{}{}{}{}{}{}O-Score}  & {\small{}{}{}{}{}{}{}{}{}{}{}{}{}{}{}{}{} 0.0140}  & {\small{}{}{}{}{}{}{}{}{}{}{}{}{}{}{}{}{} 0.0109}  & {\small{}{}{}{}{}{}{}{}{}{}{}{}{}{}{}{}{} Reject Spanning}\tabularnewline
{\small{}{}{}{}{}{}{}{}{}{}{}{}{}{}{}{}{}Return on Assets}  & {\small{}{}{}{}{}{}{}{}{}{}{}{}{}{}{}{}{} 0.0235}  & {\small{}{}{}{}{}{}{}{}{}{}{}{}{}{}{}{}{} 0.0321}  & {\small{}{}{}{}{}{}{}{}{}{}{}{}{}{}{}{}{} Spanning}\tabularnewline
{\small{}{}{}{}{}{}{}{}{}{}{}{}{}{}{}{}{}Betting against Beta}  & {\small{}{}{}{}{}{}{}{}{}{}{}{}{}{}{}{}{} 0.0404}  & {\small{}{}{}{}{}{}{}{}{}{}{}{}{}{}{}{}{} 0.0424}  & {\small{}{}{}{}{}{}{}{}{}{}{}{}{}{}{}{}{} Spanning}\tabularnewline
{\small{}{}{}{}{}{}{}{}{}{}{}{}{}{}{}{}{}Quality minus Junk}  & {\small{}{}{}{}{}{}{}{}{}{}{}{}{}{}{}{}{} 0.0304}  & {\small{}{}{}{}{}{}{}{}{}{}{}{}{}{}{}{}{} 0.0177}  & {\small{}{}{}{}{}{}{}{}{}{}{}{}{}{}{}{}{} Reject Spanning}\tabularnewline
{\small{}{}{}{}{}{}{}{}{}{}{}{}{}{}{}{}{}Size}  & {\small{}{}{}{}{}{}{}{}{}{}{}{}{}{}{}{}{} 0.0}  & {\small{}{}{}{}{}{}{}{}{}{}{}{}{}{}{}{}{} 0.0}  & {\small{}{}{}{}{}{}{}{}{}{}{}{}{}{}{}{}{} Spanning}\tabularnewline
{\small{}{}{}{}{}{}{}{}{}{}{}{}{}{}{}{}{}Growth Option}  & {\small{}{}{}{}{}{}{}{}{}{}{}{}{}{}{}{}{} 0.0029}  & {\small{}{}{}{}{}{}{}{}{}{}{}{}{}{}{}{}{} 0.0}  & {\small{}{}{}{}{}{}{}{}{}{}{}{}{}{}{}{}{} Reject Spanning}\tabularnewline
{\small{}{}{}{}{}{}{}{}{}{}{}{}{}{}{}{}{}Value (Book
to Market)}  & {\small{}{}{}{}{}{}{}{}{}{}{}{}{}{}{}{}{} 0.2045}  & {\small{}{}{}{}{}{}{}{}{}{}{}{}{}{}{}{}{} 0.1878}  & {\small{}{}{}{}{}{}{}{}{}{}{}{}{}{}{}{}{} Reject Spanning}\tabularnewline
{\small{}{}{}{}{}{}{}{}{}{}{}{}{}{}{}{}{}Idiosyncratic
Volatility}  & {\small{}{}{}{}{}{}{}{}{}{}{}{}{}{}{}{}{} 0.2386}  & {\small{}{}{}{}{}{}{}{}{}{}{}{}{}{}{}{}{} 0.0101}  & {\small{}{}{}{}{}{}{}{}{}{}{}{}{}{}{}{}{} Reject Spanning}\tabularnewline
{\small{}{}{}{}{}{}{}{}{}{}{}{}{}{}{}{}{}Profitability}  & {\small{}{}{}{}{}{}{}{}{}{}{}{}{}{}{}{}{} 0.0}  & {\small{}{}{}{}{}{}{}{}{}{}{}{}{}{}{}{}{} 0.0}  & {\small{}{}{}{}{}{}{}{}{}{}{}{}{}{}{}{}{} Spanning}\tabularnewline
\hline 
\end{tabular} 
\noindent %
\noindent\parbox[t][0.7\totalheight]{1\textwidth}{%
Entries report the 
test statistics $\rho_{T}^{\star}$
and the regression estimates $q_{T}^{BC}$  for
spanning of the Hou-Xue-Zhang (q) model with respect to each one
of the 18 market anomalies.
We reject spanning at significance level $\alpha=5\%$ if $\rho_{T}^{\star} > q_{T}^{BC}$.
The dataset spans the period
from January, 1974 to December, 2016.
} 
\end{table}

We observe that the FF-5 model spans 6 out of 18 market anomalies,
that is, Accruals, Asset Growth, Return on Assets, Size, Growth Option, and Profitability.
The M-4 model spans the same 6 market anomalies, while the q model
spans Return on Assets, Betting against Beta, Size, and Profitability. Thus, in most cases, optimal
portfolios based on the investment opportunity set that includes a
market anomaly is not spanned by the corresponding optimal portfolio
strategies based on the original factors. We also observe that Return on Assets,
Size, and Profitability are spanned by all the factor models, indicating
the robustness of these characteristics being not considered as genuine market
anomalies by prospect investors.

\subsection{Out-of-Sample Analysis}

In this section, we examine whether the inclusion of a market anomaly
in the investment opportunity set  benefits to prospect investors out-of-sample.
Although we reject the null hypothesis
of prospect spanning  in most cases for the in-sample tests, it is not known a priori whether an optimal augmented
portfolio also outperforms an optimal portfolio made of factors only in an out-of-sample
analysis. This is because by construction we form these portfolios 
at time $t$, based on the information prevailing at time $t$, while we reap
the portfolio returns over $[t,t+1]$ (next month). The
out-of-sample test is a real-time exercise mimicking the way that
a real-time investor acts.

Each time the hypothesized portfolio manager with prospect preferences
forms optimal portfolios from two separate asset universes: the first
universe consists only of factors from a factor model (FF-5, M-4,
q), the set $\mathbb{K}$. The second universe is the respective set
of factors augmented by a single trading (spread) strategy, the set
$\mathbb{L}$. Portfolio managers are assumed to solve portfolio optimization
problems, motivated by the prospect spanning framework, effectively
looking for a portfolio picked from the augmented universe $\mathbb{L}$
that prospect stochastically dominates all portfolios of the respective
factor universe $\mathbb{K}$..

The rejection of the prospect spanning hypothesis implies that there
exists at least one portfolio in $\mathbb{L}$ build from the factors (of each particular factor model) and one
market anomaly,
which is weakly prefered
to every factor portfolio in $\mathbb{K}$ by at least one S-shaped
utility function (see Definition 2). Such a portfolio is by construction
efficient w.r.t.\ $\mathbb{K}$ (see Definition 2.1 in Linton et al.\
(2014) for the SSD case which we can easily generalize to our PSD
case). The empirical version of such a portfolio is the optimal portfolio
$\lambda$ that maximizes $\rho_{T}$ for the particular sample value.
In what follows, and given this characterization, we analyze the performance
of such empirically optimal PSD portfolios through time, compared
to the performance of the optimal factor portfolios solely derived from $\mathbb{K}$
by prospect investors. 

We resort to backtesting experiments on a rolling horizon basis. The
rolling windows cover the 516 months period from 01/1974
to 12/2016. At each month, we use the data from the previous 25 years
(300 monthly observations) to calibrate the procedure. We solve the
resulting optimization problem for the prospect stochastic spanning
test and record the optimal portfolios. The clock is advanced and
we determine the realized returns of the optimal portfolios from
the actual returns of the various assets. Then we repeat the same procedure
for the next time period and we compute the ex post realized returns
over the period from 01/1999 to 12/2016 (216 months)  for both portfolios.

We compute a number of commonly used performance measures:
the average return (Mean), the standard deviation (SD) of returns, the Sharpe
ratio, the downside Sharpe ratio (D. Sharpe ratio)  of Ziemba (2005), the upside potential
and downside risk (UP) ratio of Sortino and van der Meer (1991), the
opportunity cost of Simaan (2013), and a measure of the portfolio risk-adjusted returns
net of transaction costs (Return Loss) of DeMiguel et al.\ (2009). The downside Sharpe and UP ratios are considered
to be more appropriate measures of performance than the typical Sharpe
ratio given the asymmetric return distribution of the anomalies.
For the calculation of
the opportunity cost, we use the following utility function which
satisfies the curvature of prospect theory (S-shaped): 
$U(R)= R^{\alpha}$ if $R\geq0$ or 
$-\gamma(-R)^{\beta}$ if $R<0,
$
where $\gamma$ is the coefficient of loss aversion (usually $\gamma=2.25$)
and $\alpha,\beta<1$.
We provide a short description of those performance measures in Appendix B.
In the next lines, we only detail the results of the out-of-sample tests for the Momentum
market anomaly. The latter is well  documented on diverse markets and asset classes (Asness, Moskowitz, and Pedersen (2013)). 
In the  Online Appendix, we report the performance measures
for the 5 Fama and French, the 4 Stambaugh and Yuan and the 4 Hou-Xue-Zhang
optimal factor portfolios, and the optimal augmented portfolios for all the other
market anomalies that we test.

Table \ref{t11} reports the  performance
measures for the Momentum anomaly under each factor model (Panels
A, B and C, respectively). These performance measures
supplement the evidence obtained from the in-sample analysis.
We observe that the Mean, the Sharpe ratio, downside Sharpe ratio
and UP ratio of the optimal augmented portfolio are improved with
respect to the optimal factor portfolio. Although these measures are
based on the first two moments, they support the in-sample result
that the set enlarged  with the momentum anomaly is not spanned by any factor
model. The same is true when we take into account transaction costs.
The Return Loss is always positive. The opportunity cost measure takes
into account the entire distribution of returns under a given characterization
of preferences. We observe that augmenting the factors by Momentum
increases the performance of the optimal portfolio with respect to
each factor model. The optimal weight of Momentum varies from 40\%
to 99\%, indicating the superior performance of this characteristic.

In the Online Appendix, we present analogous Tables for the other
market anomalies. Interestingly, based on the opportunity cost, enlarging
the factor set by a market anomaly increases the performance of an optimal
portfolio in 12 out of the 18 cases with respect to FF-5 factors (Composite
Equity Issue, Distress, Growth Profitability Premium, Investment to Assets, Momentum, Net Operating Assets, O-Score,
Net Stock Issues, Betting against Beta, Quality minus Junk, Value, and
Idiosyncratic Volatility), in 10 cases with respect to M-4 factors (Composite
Equity Issue, Distress, Investment to Assets, Momentum, Net Operating Assets, Net Stock Issues, Betting
against Beta, Quality minus Junk, Value, and Idiosyncratic Volatility)
and in 14 cases with respect to q factors (Accruals, Asset Growth,
Composite Equity Issue, Distress, Growth Profitability Premium, Investment to Assets, Momentum, Net Operating Assets,
O-Score, Net Stock Issues, Betting against Beta, Quality minus Junk, Size,
Value, and Idiosyncratic Volatility). For all these additional market
anomalies, we find a positive opportunity cost $\theta$. One needs
to give a positive return equal to $\theta$ to an investor who does
not include the anomalies in her portfolio so that she becomes as
happy as an investor who includes them. The computation of
the opportunity cost requires the computation of the expected utility
and hence the use of the probability density function of portfolio
returns. Thus, the calculated opportunity cost has taken into account
the higher order moments in contrast to the Sharpe ratios. Therefore,
the opportunity cost estimates provide further convincing evidence
for the diversification benefits of the inclusion of the market anomalies
given their deviation from normality.

Additionally, although the rest of the performance measures depend
mostly on the first two moments of the return distribution, they give
consistent results. The Return Loss measure that takes
into account transaction costs, is positive in all the above cases.
This reflects an increase in risk-adjusted performance (i.e., an increase
in expected return per unit of risk) and hence expands the investment
opportunities of prospect investors. The same is true for the UP ratio.
Finally, the Sharpe ratio and the downside Sharpe ratio agree that
the performance of the optimal  portfolios augmented with the above market
anomalies is improved, although  the differences are
small in some cases.

The analysis indicates that the Composite Equity Issue, Distress, Investment to Assets,
Momentum, Net Operating Assets, Net Stock Issues, Quality minus Junk, Value, and Idiosyncratic
Volatility emerge as unambiguously genuine market anomalies under
all factor sets, both in- and out-of-sample. Prospect investors
would benefit from including these characteristics in their portfolios,
expanding the investment opportunity set offered by factor portfolios.
We stress that the prospect spanning approach is particularly
robust in-sample and out-of-sample.
The remarkable consistency of in-sample
and out-of-sample results offers good incentives for adopting such an approach when
exploring instances of apparent market inefficiency.

To sum up, the in-sample spanning tests, as well as the out-of-sample
analysis given by the  performance measures, indicate
that in most cases (depending on the factor model used) the investment
universe augmented with a market anomaly dominates
the 5 Fama and French, the 4 Stambaugh and Yuan, and the 4 Hou-Xue-Zhang factors,
yielding diversification benefits and providing better investment
opportunities for investors with prospect type preferences towards
risk.

\begin{table}[H]
\caption{Performance measures. The case of the Momentum anomaly.}
\label{t11}{\small{}{}{}{}{}{}{}{}{}{}{}{}{}{}{}{}{}\centering
}%
\begin{tabular}{lcccccc}
\hline 
 & {\small{}{}{}{}{}{}{}{}{}{}{}{}{}{}{}{}Panel A}  &  & {\small{}{}{}{}{}{}{}{}{}{}{}{}{}{}{}{}Panel B}  &  & {\small{}{}{}{}{}{}{}{}{}{}{}{}{}{}{}{}Panel C}  & \tabularnewline
 & {\small{}{}{}{}{}{}{}{}{}{}{}{}{}{}{}{}{} FF-5}  & {\small{}{}{}{}{}{}{}{}{}{}{}{}{}{}{}{}{} + anom.}  & {\small{}{}{}{}{}{}{}{}{}{}{}{}{}{}{}{}{} M-4}  & {\small{}{}{}{}{}{}{}{}{}{}{}{}{}{}{}{}{} + anom.}  & {\small{}{}{}{}{}{}{}{}{}{}{}{}{}{}{}{}{} q}  & {\small{}{}{}{}{}{}{}{}{}{}{}{}{}{}{}{}{} + anom.}\tabularnewline
\hline 
{\small{}{}{}{}{}{}{}{}{}{}{}{}{}{}{}{}{}Mean}  & {\small{}{}{}{}{}{}{}{}{}{}{}{}{}{}{}{}{}0.0056}  & {\small{}{}{}{}{}{}{}{}{}{}{}{}{}{}{}{}{}0.0062}  & {\small{}{}{}{}{}{}{}{}{}{}{}{}{}{}{}{}{}0.0044}  & {\small{}{}{}{}{}{}{}{}{}{}{}{}{}{}{}{}{}0.0048}  & {\small{}{}{}{}{}{}{}{}{}{}{}{}{}{}{}{}{}0.0073}  & {\small{}{}{}{}{}{}{}{}{}{}{}{}{}{}{}{}{}0.0072}\tabularnewline
{\small{}{}{}{}{}{}{}{}{}{}{}{}{}{}{}{}{}SD}  & {\small{}{}{}{}{}{}{}{}{}{}{}{}{}{}{}{}{}0.0358}  & {\small{}{}{}{}{}{}{}{}{}{}{}{}{}{}{}{}{}0.0370}  & {\small{}{}{}{}{}{}{}{}{}{}{}{}{}{}{}{}{}0.0388}  & {\small{}{}{}{}{}{}{}{}{}{}{}{}{}{}{}{}{}0.0409}  & {\small{}{}{}{}{}{}{}{}{}{}{}{}{}{}{}{}{}0.0808}  & {\small{}{}{}{}{}{}{}{}{}{}{}{}{}{}{}{}{}0.0385}\tabularnewline
{\small{}{}{}{}{}{}{}{}{}{}{}{}{}{}{}{}{}Sharpe ratio}  & {\small{}{}{}{}{}{}{}{}{}{}{}{}{}{}{}{}{}0.1507}  & {\small{}{}{}{}{}{}{}{}{}{}{}{}{}{}{}{}{}0.1604}  & {\small{}{}{}{}{}{}{}{}{}{}{}{}{}{}{}{}{}0.1063}  & {\small{}{}{}{}{}{}{}{}{}{}{}{}{}{}{}{}{}0.1117}  & {\small{}{}{}{}{}{}{}{}{}{}{}{}{}{}{}{}{}0.0879}  & {\small{}{}{}{}{}{}{}{}{}{}{}{}{}{}{}{}{}0.1814}\tabularnewline
{\small{}{}{}{}{}{}{}{}{}{}{}{}{}{}{}{}{}D. Sharpe
ratio}  & {\small{}{}{}{}{}{}{}{}{}{}{}{}{}{}{}{}{}0.1622}  & {\small{}{}{}{}{}{}{}{}{}{}{}{}{}{}{}{}{}0.1706}  & {\small{}{}{}{}{}{}{}{}{}{}{}{}{}{}{}{}{}0.1078}  & {\small{}{}{}{}{}{}{}{}{}{}{}{}{}{}{}{}{}0.1108}  & {\small{}{}{}{}{}{}{}{}{}{}{}{}{}{}{}{}{}0.0868}  & {\small{}{}{}{}{}{}{}{}{}{}{}{}{}{}{}{}{}0.1995}\tabularnewline
{\small{}{}{}{}{}{}{}{}{}{}{}{}{}{}{}{}{}UP ratio}  & {\small{}{}{}{}{}{}{}{}{}{}{}{}{}{}{}{}{}0.6401}  & {\small{}{}{}{}{}{}{}{}{}{}{}{}{}{}{}{}{}0.6693}  & {\small{}{}{}{}{}{}{}{}{}{}{}{}{}{}{}{}{}0.5646}  & {\small{}{}{}{}{}{}{}{}{}{}{}{}{}{}{}{}{}0.5853}  & {\small{}{}{}{}{}{}{}{}{}{}{}{}{}{}{}{}{}0.5348}  & {\small{}{}{}{}{}{}{}{}{}{}{}{}{}{}{}{}{}0.6769}\tabularnewline
{\small{}{}{}{}{}{}{}{}{}{}{}{}{}{}{}{}{}Return Loss}  &  & {\small{}{}{}{}{}{}{}{}{}{}{}{}{}{}{}{}{}0.0351\%}  &  & {\small{}{}{}{}{}{}{}{}{}{}{}{}{}{}{}{}{}0.0205\%}  &  & {\small{}{}{}{}{}{}{}{}{}{}{}{}{}{}{}{}{}0.3723\%}\tabularnewline
\hline 
{\small{}{}{}{}{}{}{}{}{}{}{}{}{}{}{}{}{}Opportunity
Cost}  &  &  &  &  &  & \tabularnewline
{\small{}{}{}{}{}{}{}{}{}{}{}{}{}{}{} $\alpha=\beta=0.2$}  &  & {\small{}{}{}{}{}{}{}{}{}{}{}{}{}{}{}{}{}0.0416\%}  &  & {\small{}{}{}{}{}{}{}{}{}{}{}{}{}{}{}{}{}0.1446\%}  &  & {\small{}{}{}{}{}{}{}{}{}{}{}{}{}{}{}{}{}0.4338\%}\tabularnewline
{\small{}{}{}{}{}{}{}{}{}{}{}{}{}{}{} $\alpha=\beta=0.4$}  &  & {\small{}{}{}{}{}{}{}{}{}{}{}{}{}{}{}{}{}0.0210\%}  &  & {\small{}{}{}{}{}{}{}{}{}{}{}{}{}{}{}{}{}0.0129\%}  &  & {\small{}{}{}{}{}{}{}{}{}{}{}{}{}{}{}{}{}0.4093\%}\tabularnewline
{\small{}{}{}{}{}{}{}{}{}{}{}{}{}{}{} $\alpha=\beta=0.6$}  &  & {\small{}{}{}{}{}{}{}{}{}{}{}{}{}{}{}{}{}0.0129\%}  &  & {\small{}{}{}{}{}{}{}{}{}{}{}{}{}{}{}{}{}0.0152\%}  &  & {\small{}{}{}{}{}{}{}{}{}{}{}{}{}{}{}{}{} 0.3229 \%}\tabularnewline
\hline 
\end{tabular}
\begin{centering}
\begin{tabular}{lccccc}
\multicolumn{6}{l}{Descriptive statistics of the weight allocation of the optimal portfolios}\tabularnewline
\hline 
 &  & Mean  & Std. Dev.  & Skewness  & Kurtosis\tabularnewline
\hline 
FF-5 Factors  & Market  & 0.5955  & 0.1507  & -3.3717  & 10.9074\tabularnewline
 & SMB  & 0.0  & 0.0  & -  & -\tabularnewline
 & HML  & 0.0  & 0.0  & -  & -\tabularnewline
 & RMW  & 0.0  & 0.0  & -  & -\tabularnewline
 & CMA  & 0.0  & 0.0  & -  & -\tabularnewline
 & Momentum  & 0.4045  & 0.1507  & 3.3717  & 10.9074\tabularnewline
\hline 
M-4 Factors  & Market  & 0.5331  & 0.2255  & -1.6812  & 1.5383\tabularnewline
 & SMB  & 0.0  & 0.0  & -  & -\tabularnewline
 & MGMT1  & 0.0020  & 0.0113  & 7.4184  & 59.9621\tabularnewline
 & PERf1  & 0.0  & 0.0  & -  & -\tabularnewline
 & Momentum  & 0.4648  & 0.2273  & 1.6464  & 1.4817\tabularnewline
\hline 
q Factors  & Market  & 0.0028  & 0.0411  & 14.6969  & 216.000\tabularnewline
 & ME  & 0.0  & 0.0  & -  & -\tabularnewline
 & IA  & 0.0  & 0.0  & -  & -\tabularnewline
 & ROE  & 0.0  & 0.0  & -  & -\tabularnewline
 & Momentum  & 0.9972  & 0.0411  & -14.6969  & 216\tabularnewline
\hline 
\end{tabular}
\par\end{centering}
\noindent %
\noindent\parbox[t][0.7\totalheight]{1\textwidth}{%
Entries report the performance measures (Mean, Standard Deviation,
Sharpe ratio, Downside Sharpe ratio, UP ratio, Returns Loss and Opportunity
Cost) for the factor optimal portfolios, as well as the augmented
with the Momentum optimal portfolio. The dataset spans the period
from January, 1999 to December, 2016. Panel A report measures for
the case of the FF-5 factors. Panel B for the case of the M-4 factors,
while panel C for the case of the q factors. In the second half, the
Table exhibits the descriptive statistics of the weight allocation
of the optimal augmented portfolios.%
} 
\end{table}

\section{Conclusions}

In this paper, we develop and implement methods for determining whether
introducing new securities or relaxing investment constraints improves
the investment opportunity set for prospect investors. We develop
a testing procedure for prospect spanning  for two nested portfolio
sets based on subsampling and standard LP.

In the empirics, we apply the prospect spanning framework
to asset prices in which investors evaluate risk according to prospect
theory and examine its ability to explain 18 well-known stock market
anomalies. The setting deploys prospect theory in a fully nonparametric way. We find that of
the strategies considered, many expand the opportunity set of the
prospect investors, thus have real economic value for them.

Most importantly, we show that the prospect spanning approach is particularly
robust between in-sample and out-of-sample applications. The paper
contributes to a current strand of literature aiming to reevaluate
published anomalies and discern those with real economic content for
prospect investors. From a practitioner perspective, this robust framework
for establishing investment  opportunities for prospect investors can
be of real value, especially in the case of quantitative investment
funds that combine talent, capital and computational power to the
purpose of exploiting the existing anomalies and discovering new ones.

$ $

{\bf APPENDIX A: Description of Stock Market Anomalies}
\\

Below we provide the origin and a short description of the 18
market anomalies used in the empirical application.

1. Accruals: Sloan (1996) argues that investors tend to overestimate
in their earnings expectations the persistence of the earnings' component
that is due to accruals. As a result, firms with low accruals earn
on average abnormally higher returns than firms with high accruals.

2. Asset Growth: Cooper, Gulen, and Schill (2008) maintain that investors
tend to overreact positively right after asset expansions. According
to the authors, this behavior causes firms with high growth in their
total assets to exhibit relatively lower returns over the subsequent
fiscal years.

3. Composite Equity Issues: Daniel and Titman (2006) base their analysis
on a measure of equity issuance that they devised finding that equity
issuers tend to underperform non-issuer firms.

4. Distress: Campbell, Hilscher, and Szilagyi (2008) find that firms
with high default probability tend to exhibit lower subsequent returns.
This pattern is counter-intuitive in the context of rational asset
pricing, given that according to the standard models high risk entails
high expected return and vice versa.

5. Gross Profitability Premium: Novy-Marx (2013) argues that gross
profit is the most objective profitability metric. As a result, firms
with the strongest gross profit have on average higher returns than
the less profitable ones.

6. Investment to Assets: Titman, Wei, and Xie (2004) argue that investors
are put off by empire-building managers who over-invest. For this
reason, firms showing a significant increase in gross property, plant,
equipment or inventories tend to underperform the market.

7. Momentum: Momentum (Jegadeesh and Titman (1993)) is perhaps the
most cited anomaly in asset pricing. Since Carhart factor model
(1997), it has been included in various reduced-form models of the
SDF as a factor. The momentum effect is attributed to sentiment and
describes the pattern of ``winner'' stocks gaining higher subsequent
returns and ``loser'' stocks relatively lower.

8. Net Operating Assets: Hirshleifer et al.\ (2004) suggest that investors
often neglect information about cash profitability and focus instead
on accounting profitability. Because of this bias, firms with high
net operating assets (measured as the cumulative difference between
operating income and free cash flow) get to have negative long-run
stock returns.

9. Net Stock Issues: Ritter (1991) and Loughran and Ritter (1995)
indicate that equity issuers underperform non-issuers with similar
characteristics. Fama and French (2008) demonstrate that net stock
issues are negatively correlated with subsequent returns.

10. O-Score: This anomaly coincides with the distress anomaly we mentioned
earlier. In this case, the spread portfolios are constructed from
stock ranking based on the O-score (Ohlson (1980)) to measure distress
likelihood.

11. Return on Assets: Chen, Novy-Marx, and Zhang
(2010) associate high past return
on assets with abnormally high subsequent returns. Return on assets
is measured as the ratio of quarterly earnings to last quarter's assets.

12. Betting against Beta: Black, Jensen and Scholes (1972) showed
that low (high) beta stocks have consistently positive (negative)
risk-adjusted returns. Frazzini and Pedersen (2014) propose an investment
strategy (``betting-against-beta'' (BAB)) that exploits this anomaly
by buying low-beta stocks and shorting high-beta stocks. Because of
its robustness, this anomaly is currently one of the most widely examined
APT violations.

13. Quality minus Junk: Asness, Frazzini and Pedersen (2013) show
that high-quality stocks (safe, profitable, growing, and well managed)
exhibit high risk-adjusted returns. The authors attribute this pattern
to mispricing.

14. Size: The market capitalization. is computed as  the log of the product of price per
share and number of shares outstanding, computed at the end of the
previous month.

15. Growth Option: Growth Option measure represents the residual future-oriented
firm growth potential. This future (yet-to-be exercised) growth option
measure is calculated as the \% of a firm's market value (V)
arising from future-oriented growth opportunities (PVGO/V). It is
inferred by subtracting from the current market value of the firm
(V) the perpetual discounted stream of expected operating cash flows
under a no-further growth policy (see, e.g., Kester (1984), Anderson
and Garcia-Feijoo (2006), Berk, Green, and Naik (1999)).

16. Value (Book to market): The log of book value of equity scaled
by market value of equity, computed following Fama and French (1992)
and Fama and French (2008); firms with negative book value are excluded
from the analysis.

17. Idiosyncratic Volatility: Standard deviation of the residuals
from a firm-level regression of daily stock returns on the daily Fama-French
three factors using data from the past month. See Ang et al. (2006).

18. Profitability.: It is measured as revenue minus cost of goods sold at
time t, divided by assets at time t-1. Stocks with high profitability
ratios tend to outperform on a risk-adjusted basis (Novy-Marx
(2013), Novy-Marx and Velikov (2015)). Recent
research suggests that profitability is one of the stock return anomalies
that has the largest economic significance (see Novy-Marx (2013)).
\\

{\bf APPENDIX B: Description of Performance Measures}
\\

For the downside Sharpe ratio, first we need to calculate the downside
variance (or more precisely the downside risk), 
$
\sigma_{P_{-}}^{2}=\frac{\sum_{t=1}^{T}(x_{t}-\bar{x})_{-}^{2}}{T-1},
$
where the benchmark $\bar{x}$ is zero, and the $x_{t}$ taken are
those returns of portfolio $P$ at month $t$ below $\bar{x}$, i.e.,
those $t$ of the $T$ months with losses. To get the total variance,
we use twice the downside variance namely $2\sigma_{P_{-}}^{2}$ so
that the downside Sharpe ratio is, 
$
S_{P}=\frac{\bar{R}_{p}-\bar{R}_{f}}{\sqrt{2}\sigma_{P-}},
$
where $\bar{R}_{p}$ is the average period return of portfolio $P$
and $\bar{R_{f}}$ is the average risk free rate. The UP ratio compares
the upside potential to the shortfall risk over a specific target
(benchmark) and is computed as follows. Let $R_{t}$ be the realized
monthly return of portfolio $P$ for $t=1,...,T$ of the backtesting
period, where $T=216$ is the number of experiments performed and
let $\rho_{t}$ be respectively the return of the benchmark (risk
free rate) for the same period. Then, we have, 
$
\textrm{UP ratio}=\frac{\frac{1}{K}\sum_{t=1}^{K}\max[0,R_{t}-\rho_{t}]}{\sqrt{{\frac{1}{K}\sum_{t=1}^{K}(\max[0,\rho_{t}-R_{t}])^{2}}}}.
$
It is obvious that the numerator of the above ratio is the average
excess return over the benchmark and so reflects upside potential.
In the same way, the denominator measures downside risk, i.e. shortfall
risk over the benchmark.

Next, we use the concept of opportunity cost presented in Simaan (2013)
to analyse the economic significance of the performance difference
of the two optimal portfolios. Let $R_{Aug}$ and $R_{F}$ be the
realized returns of the optimal augmented and the optimal factors
portfolios, respectively. Then, the opportunity cost $\theta$ is
defined as the return that needs to be added to (or subtracted from)
the optimal factors portfolio return $R_{F}$, so that the investor
is indifferent (in utility terms) between the strategies imposed by
the two different investment opportunity sets, i.e.,
$
E[U(1+R_{F}+\theta)]=E[U(1+R_{Aug})].
$

A positive (negative) opportunity cost implies that the investor is
better (worse) off if the investment opportunity set allows for the
market anomaly factor prospect type investing. The opportunity cost
takes into account the entire probability density function of asset
returns and hence it is suitable to evaluate strategies even when
the asset return distribution is not normal. For the calculation of
the opportunity cost, we use the following utility function which
satisfies the curvature of prospect theory (S-shaped): 
$U(R)= R^{\alpha}$ if $R\geq0$ or 
$-\gamma(-R)^{\beta}$ if $R<0,
$
where $\gamma$ is the coefficient of loss aversion (usually $\gamma=2.25$)
and $\alpha,\beta<1$.

Finally, we evaluate the performance of the two portfolios under the
risk-adjusted (net of transaction costs) returns measure, proposed
by DeMiguel et al.\ (2009) which indicates the way that the proportional
transaction cost, generated by the portfolio turnover, affects the
portfolio returns. Let $trc$ be the proportional transaction cost,
and $R_{P,t+1}$ the realized return of portfolio $P$ at time $t+1$.
The change in the net of transaction cost wealth $NW_{P}$ of portfolio
$P$ through time is,
$
NW_{P,t+1}=NW_{P,t}(1+R_{P,t+1})[1-trc\times\sum_{i=1}^{N}(|w_{P,i,t+1}-w_{P,i,t}|).
$
The portfolio return, net of transaction costs is defined as
$
RTC_{P,t+1}=\frac{NW_{P,t+1}}{NW_{P,t}}-1.
$
Let $\mu_{F}$ and $\mu_{Aug}$ be the out-of-sample mean of monthly
$RTC$ factros and the Augmented optimal portfolio, respectively,
and $\sigma_{F}$ and $\sigma_{Aug}$ be the corresponding standard
deviations. Then, the return-loss measure is,
$
R_{Loss}=\frac{\mu_{Aug}}{\sigma_{Aug}}\times\sigma_{F}-\mu_{F},
$
i.e., the additional return needed so that the factors performs equally
well with the optimal augmented with the market anomaly portfolio.
We follow the literature and use 35 bps for the transaction cost.

\end{document}